\documentclass[twocolumn,numberedappendix]{openjournal}
\usepackage{amsmath}
\usepackage{xcolor}
\usepackage{textgreek}
\usepackage[utf8]{inputenc}
\usepackage[english]{babel}

\usepackage{hyperref}
\hypersetup{
unicode,
colorlinks=true,
linkcolor=linkcolor,
citecolor=linkcolor,
filecolor=linkcolor,
urlcolor=linkcolor,
}
\usepackage{color,colortbl}
\definecolor{linkcolor}{rgb}{0.0,0.3,0.5}
\usepackage{tensind}
\tensordelimiter{?}
\DeclareGraphicsExtensions{.bmp,.png,.jpg,.pdf}
\usepackage{verbatim}
\usepackage[normalem]{ulem}
\usepackage{orcidlink}
\usepackage{soul}

\graphicspath{ {./figs/} }

\newcommand{\Mstar}{M_\star}
\newcommand{\Mhalo}{M_{\rm h}}
\newcommand{\Mgas}{M_{\rm g}}
\newcommand{\Msun}{\,{\rm M}_\odot}
\newcommand{\tdyn}{t_{\rm dyn}}

\newcommand{\epsfb}{\varepsilon_{\rm fb}}
\newcommand{\epsff}{\varepsilon_{\rm ff}}
\newcommand{\Zs}{Z_\star}
\newcommand{\Vesc}{V_{\rm esc}}

\begin{document}

\title{Discrete Supernovae and the Limits of Continuous Models in Low-Mass Dark Matter Halos}

\author{Chris Power\altaffilmark{1}\orcidlink{0000-0002-4003-0904} and Sreedhar Balu\altaffilmark{2}\orcidlink{0000-0002-5281-5151}}
\thanks{Corresponding author: chris.power@uwa.edu.au}

\affiliation{$^1$International Centre for Radio Astronomy Research, University of Western Australia, 35 Stirling Highway, Crawley, Western Australia 6009, Australia}

\affiliation{$^2$Facultad de F\'isicas, Multidisciplinary Unit for Energy Science, Universidad de Sevilla, 41012, Seville, Spain}

\begin{abstract}
Many galaxy formation models represent supernova feedback by a continuous energy injection rate averaged over the stellar initial mass function (IMF). This approximation can fail in low-mass dark matter halos, where single, discrete supernovae can unbind more gas than the reservoir holds. We investigate this using a one-zone model that characterises continuous and discrete supernova feedback at redshifts $z\geq5$. Both feedback modes inject the same expected energy per unit stellar mass formed. In the discrete model, individual supernovae are resolved and energy is injected instantaneously; in the continuous model, supernovae are IMF-averaged and energy injection is spread over a finite time interval. Once stars form, a supernova can unbind 1.5--2.7 times the gas present. Continuous feedback begins removing gas after $\sim 4$\,Myr, whereas the first discrete supernova occurs after $\sim 10$\,Myr, which produces a stellar mass excess before feedback takes effect. The excess persists because the reservoir responds nonlinearly: an event removes at most the gas present, and star formation halts once it is empty. It depends strongly on halo mass: from 135 per cent at $10^7\,\Msun$, to 15 per cent at $10^8\,\Msun$, to negligible at $10^9\,\Msun$ by $z=5$. Fixing the number and timing of supernovae does not remove the stellar excess, which indicates that it is not caused by stochastic IMF sampling. Instead, it vanishes when the same energy is divided among more, smaller events. Matching the energy released per unit stellar mass therefore does not guarantee equivalent feedback in halos that retain gas until the first supernova.
\end{abstract}

\section{Introduction}
\label{sec:intro}

Galaxy formation models commonly represent supernova feedback as a continuous process, removing gas at a rate proportional to the star formation rate. The proportionality is determined by the number of supernovae per unit stellar mass formed, averaged over the stellar initial mass function (IMF) \citep[e.g.][]{Cole2000,SpringelHernquist2003,Croton2006}. This approximation replaces individual supernovae (hereafter events) with their mean effect on the gas reservoir (hereafter reservoir). It is reasonable when individual events produce small perturbations to the gas on which they act \citep{DekelSilk1986,Efstathiou1992}, but becomes questionable in low-mass halos, where a single supernova can unbind most or all of the gas in the reservoir \citep[e.g.][]{Bromm2003,KitayamaYoshida2005,Whalen2008}. An analogous resolution dependence is now being encountered directly in star-by-star hydrodynamical simulations (\S\ref{sec:discussion}). \S\ref{sec:model} formalises one specific implementation of this averaged treatment, which we call the \emph{baseline continuous model} throughout; we distinguish it explicitly from the generic IMF-averaged prescriptions discussed in this Introduction, and from the alternative continuous models tested in Appendix~\ref{sec:appregime}.

In this regime, the response to feedback depends not only on the total energy released, but also on when individual events occur and how much gas the reservoir holds at the time. Supernovae from a stellar population explode over a range of progenitor lifetimes, from a few to several tens of Myr \citep{Georgy2013}. An event cannot remove more gas than the reservoir contains: energy in excess of that required to expel the gas present may be lost, while subsequent accretion replenishes the reservoir between events. In a continuous prescription, feedback is applied in smaller increments as gas is present, whereas a discrete event can release more energy than is required to remove the reservoir. Consequently, the two treatments need not produce the same integrated gas removal or stellar mass, even when they have the same energy normalisation per unit stellar mass formed. The reason is that gas removal saturates at the gas present: averaging the energy before applying this limit does not give the same result as applying it to each event and then averaging the outcomes.

Some galaxy formation models account for the finite lifetimes of massive stars by delaying feedback relative to star formation \citep[e.g.][]{Mutch2016,Henriques2020,Hutter2021,MenonPower2024,MenonBaluPower2026}. Delayed feedback can produce oscillations in the gas mass and star formation rate \citep{FaucherGiguere2018,MenonPower2024}, but does not necessarily capture the response to individual supernovae. Other studies have examined the effects of discrete energy injection and stochastic IMF sampling. These approaches have found effects ranging from negligible to substantial, including changes in wind driving, stellar-mass regulation and chemical enrichment \citep{Su2018,Applebaum2020,JeonGo2026}. The physical reason to expect an effect is that a small stellar population produces relatively few supernovae: a few thousand solar masses of stars yields only a few tens of supernovae. Individual events can therefore represent substantial perturbations to the reservoir, particularly in low-mass halos, where the energy coupled to the gas by one supernova can exceed that required to unbind it. The number and timing of supernovae influence when and how much gas is removed, while stochastic IMF sampling also affects the energy and metals returned by the stellar population. The importance of these effects depends on the reservoir relative to the mass a supernova can unbind, and on the other physical processes included in a simulation. Consequently, it remains difficult to establish the relative importance of discrete energy injection and IMF sampling across existing calculations. Throughout this paper we distinguish discrete energy injection from stochastic IMF sampling: the former, not the latter, is sufficient to produce the effect we examine (\S\ref{sec:size}).

Stochastic variations in star formation provide another way to depart from a continuous description. Models that introduce correlated fluctuations in the star formation rate while preserving its mean can produce substantial variations in galaxy properties \citep{PanKravtsov2023,KravtsovBelokurov2024}. However, when outflows remain tied to the mean star formation rate, these prescriptions do not capture the response of a finite reservoir to individual supernovae. Stochastic star formation and discrete feedback are therefore distinct physical effects. The response of low-mass halos is also affected by processes such as photoionisation, gas re-accretion, metal enrichment and the structure of the interstellar medium \citep{KellerKruijssen2022,Smith2021,GrudicHopkins2019,Grudic2023,Hirai2021,Sanati2023,Andersson2026}. These processes influence whether gas survives and how it responds to feedback, but make it difficult to isolate the effect of applying supernova feedback as individual events rather than as a continuous rate.

The goal of this paper is to isolate this distinction. We ask when replacing individual supernovae with their IMF-averaged continuum limit changes the mean stellar mass assembled by a halo, and what drives that difference. In particular, we distinguish the effects of the finite reservoir from those of the delay before the first supernova, and examine how the difference depends on halo mass, feedback coupling and the treatment of excess energy.

We address these questions using \textsc{Ashvini} \citep{MenonPower2024,MenonBaluPower2026}, a lightweight semi-analytic model of galaxy formation and evolution. We model cosmological gas accretion that tracks halo assembly histories (via merger trees); star formation from a single reservoir; and gas removal from this reservoir by supernovae. 
We compare feedback from discrete events, each limited to ejecting the gas in the reservoir at the time of the event, with a continuous model that releases the same energy per unit stellar mass formed. By fixing the number and timing of supernovae, we distinguish the effects of discrete energy injection from those of stochastic IMF sampling. We examine how the difference depends on halo mass, evolutionary time, feedback coupling and the energy released by individual events. In particular, we separate the effects of the finite reservoir from those of the delay before the first supernova.

The paper is organised as follows. \S\ref{sec:model} describes the model. \S\ref{sec:physics} presents the results, establishing the halo mass dependence and physical origin of the difference between discrete and continuous feedback. \S\ref{sec:discussion} discusses the implications and the limitations of the model. \S\ref{sec:conclusions} summarises the main results of the paper. Additional details of the modelling are presented in the Appendices. 

% ============================================================
\section{Model}
\label{sec:model}
% ============================================================

\subsection{Physical assumptions}

We consider a dark matter halo at a time $t$ with mass $\Mhalo(t)$. The halo accumulates a reservoir of gas of mass $\Mgas$ via cosmological accretion, which tracks the growth of the main branch of its merger tree (\S\ref{sec:implementation}). The rate at which $\Mgas$ grows via cosmological accretion is given by,
\begin{equation}
\dot M_{\rm acc}=f_{\rm b}\,f_{\rm ret}\,\zeta\,\max(\dot M_{\rm halo},0),
\label{eq:acc}
\end{equation}
where $f_{\rm b}=0.157$ is the cosmic baryon fraction, $f_{\rm ret}=1$, and $\zeta$ parameterises the suppression of the growth as a consequence of cosmological reionisation at $z_{\rm rei}=8$ \citep{Okamoto2008,MenonPower2024}. Our standard assumption is that the accreted gas is pristine (i.e. has zero metallicity). Stars form from this gas at a fixed efficiency per dynamical time,
\begin{equation}
\dot M_\star=\epsff\,\frac{\Mgas}{\tdyn},\qquad \tdyn=\left(200\,G\rho_{\rm crit}\right)^{-1/2},
\label{eq:sf}
\end{equation}
with $\epsff=0.05$ and $\rho_{\rm crit}(z)$ the critical density. Star formation can proceed once the halo first exceeds a threshold mass for cooling,
\begin{equation}
M_{\rm thr}(z)=
\begin{cases}
7\times10^{5}\,\Msun, & z>z_{\rm rei},\\[4pt]
\max\!\left[7\times10^{5}\,\Msun,\ M_{\rm cool}(z)\right], & z\leq z_{\rm rei}.
\end{cases}
\label{eq:cross}
\end{equation}
The value of $7\times10^{5}\,\Msun$ corresponds to the threshold mass found by \citet{Lenoble2024} for cold, dense, collapsing gas, and $M_{\rm cool}(z)$ is the halo mass with a virial temperature of $10^{4}$\,K, the mass needed for atomic cooling. We measure time from the moment the halo reaches this threshold, when star formation begins. The initial reservoir is the filtered baryon content of the halo at that time.

The stellar mass formed produces supernovae after the lifetimes of their progenitors. In a discrete run, stellar mass $\Delta\Mstar$ formed at time $t_{\rm f}$ produces a Poisson number of supernovae, each exploding after the lifetime of its progenitor,
\begin{equation}
N_{\rm SN}\sim{\rm Poisson}\left(\nu_{\rm SN}\,\Delta\Mstar\right),\qquad t_{\rm SN}=t_{\rm f}+\tau(m),
\label{eq:sn}
\end{equation}
with $\nu_{\rm SN}=1.18\times10^{-2}\,\Msun^{-1}$ for a Chabrier IMF \citep{Chabrier2003} over $8$--$100\,\Msun$, progenitor masses $m$ drawn from the IMF, and the lifetime $\tau(m)$ from the rotating $Z=0.002$ tracks of \citet{Georgy2013}. Each supernova removes from the reservoir the smaller of the gas present and the mass its energy can unbind,
\begin{equation}
\Delta M_{\rm ej}=\min\!\left(\Mgas,\;m_1\right),\qquad m_1=\frac{2\,\epsfb E_{\rm SN}}{\Vesc^{2}},
\label{eq:dmej}
\end{equation}
with $E_{\rm SN}=10^{51}$\,erg, a coupling $\epsfb=0.1$, and the escape velocity at the time of the event
\begin{equation}
\Vesc=\sqrt2\,V_{\rm c},\qquad V_{\rm c}^{2}=\frac{G\Mhalo}{r_{\rm vir}},
\label{eq:vesc}
\end{equation}
where $r_{\rm vir}$ encloses a mean density of $200\rho_{\rm crit}$. An event removes at most the gas present in the reservoir, and any energy that finds no gas to unbind is lost. Gas removed in this way leaves the halo and does not return, except in the mass-return variant described below. The reservoir is a single zone, and apart from the variants below it has no memory of earlier events beyond the gas it holds. Metals released by each supernova mix instantly into the reservoir, the gas removed by an event carries away its share of the metals, and new stars form with the metallicity of the reservoir. Each supernova releases
\begin{equation}
M_{Z,{\rm SN}}=\frac{Y_Z}{\nu_{\rm SN}},\qquad Y_Z=0.06,
\label{eq:metals}
\end{equation}
so that $0.06\,\Msun$ of metals are produced per unit stellar mass formed. 

We call runs in which supernovae are individual events \emph{discrete runs}; in the fiducial runs their number and timing follow Equation~\ref{eq:sn}, and in some tests they are fixed (\S\ref{sec:size}). Because the number of supernovae is Poisson distributed, the discrete and continuous models inject the same energy per unit stellar mass formed in expectation; the energy released in an individual discrete run fluctuates about this value. 

We compare the discrete runs with a continuous feedback model, hereafter the baseline continuous model (or, where unambiguous, simply the continuous model). It has the same stellar mass formed, lifetimes and energy input, but the energy is released in many small increments, as the IMF-averaged supernova rate implies, in place of individual events. The IMF over $8$--$100\,\Msun$ is divided into $N_{\rm bin}=200$ bins of equal probability. Stellar mass formed in each 1\,Myr is released as one episode of $N_{\rm bin}$ increments, at the formation time plus the lifetime of each bin, every increment carrying $1/N_{\rm bin}$ of the supernovae expected, so that the supernova rate follows the star formation rate:
\begin{equation}
\Delta M_{{\rm ej},k}=\min\!\left(\Mgas,\;w_k\,m_1\right),\qquad w_k=\frac{\nu_{\rm SN}\,\Delta\Mstar}{K}.
\label{eq:control}
\end{equation}
The continuous model is the limit of the discrete model as the energy per event tends to zero, but it is not the mean of the discrete runs. The reservoir responds nonlinearly; an event can remove at most the gas present (Equation~\ref{eq:dmej}), and star formation stops when the reservoir is empty. Averaging the energy first and then applying these limits therefore gives a different stellar mass from applying them to each event and then averaging. We measure the offset in the mean stellar mass,
\begin{equation}
\delta_\star=\frac{\langle\Mstar\rangle}{\Mstar^{\rm cont}}-1,
\label{eq:offset}
\end{equation}
where $\langle\Mstar\rangle$ is the mean stellar mass over the realisations of the discrete model for the same halo history, each with a different set of random numbers for the numbers and delays of supernovae. For an ensemble of trees we take the offset of each tree and quote the median over trees, with the 16--84 per cent range from bootstrapping the trees. 
% The offset is dimensionless: the figures show it as a fraction and the text quotes it in per cent, so $\delta_\star=1$ is an excess of 100 per cent. 
The offset in stellar metallicity is defined in the same way, $\delta_Z=\langle\Zs\rangle/\Zs^{\rm cont}-1$, with $\Zs=M_{Z,\star}/\Mstar$.

We vary the baseline treatment of an event in three ways: surplus energy is retained for later inflow (\emph{energy retention}), a fraction of the ejected gas returns to the reservoir (\emph{mass return}), and the coupling differs between the discrete runs and the continuous model (\emph{varied coupling}). In \emph{energy retention}, the part of an event's energy that finds no gas,
\begin{equation}
E_{\rm s}=\tfrac12\,\left(m_1-\Delta M_{\rm ej}\right)\Vesc^{2},
\label{eq:surplus}
\end{equation}
is kept in the halo as a store $E_{\rm b}$ and used to remove gas that arrives later, as if the event had left a hot halo that unbinds fresh inflow. In each step the store removes
\begin{equation}
\Delta m_{\rm b}=\min\!\left(\dot M_{\rm acc}\,\Delta t,\ \frac{2E_{\rm b}}{\Vesc^{2}}\right)
\label{eq:bank}
\end{equation}
from the inflow, at a cost of $\tfrac12\,\Delta m_{\rm b}\,\Vesc^{2}$ in $E_{\rm b}$, and between events it decays as $E_{\rm b}\propto e^{-t/\tau_{\rm b}}$, with $\tau_{\rm b}=100$\,Myr or $\infty$; the second is an upper bound, since the energy then suffers no radiative loss. In \emph{mass return}, a fraction $f_{\rm rec}$ of each event's ejecta, with its metals, returns to the reservoir after the lag
\begin{equation}
\tau_{\rm rec}=\kappa_{\rm rec}\,\frac{r_{\rm vir}}{\Vesc},
\label{eq:return}
\end{equation}
with $\kappa_{\rm rec}=1$. In \emph{varied coupling}, $\epsfb$ is changed, or differs between the sampled runs and the continuous model.

\subsection{Numerical implementation}
\label{sec:implementation}

Merger trees are generated with the \citet{Parkinson2008} algorithm implemented within the \textsc{foraois} \citep{PowerBalu2026} code, with a halo-mass resolution of $10^4\,\Msun/h$. We follow the main branch, which reaches a final specified target mass at $z=5$. The reservoir model of \S\ref{sec:model}, including the discrete and continuous feedback treatments, is implemented within \textsc{Ashvini} \citep{MenonPower2024,MenonBaluPower2026}, driven by these trees. The reservoir is advanced in substeps of at most 1\,Myr, split at every pending event, with the star formation equation (Equation~\ref{eq:sf}) solved exactly in each substep (Appendix~\ref{sec:appnum}).

The fiducial ensemble has 100 trees reaching each of the masses $5\times10^6$, $10^7$, $3\times10^7$, $10^8$ and $10^9\,\Msun$ at $z=5$, with 50 realisations of the supernova sampling per tree. At $3\times10^9$, $10^{10}$ and $3\times10^{10}\,\Msun$ we use three realisations per tree.  In each discrete run, the random sampling determines the number of supernovae produced by each episode of star formation and the delay before each explodes (Equation~\ref{eq:sn}). These quantities are drawn independently for every run in every tree, so no two runs share the same sampling. The merger trees themselves are generated separately. The spread between runs of a given tree therefore measures variation due to supernova sampling alone; averaging over trees does not introduce correlated sampling noise between trees.

% ============================================================
\section{Results}
\label{sec:physics}
% ============================================================

\subsection{The effect on stellar mass}
\label{sec:mean}

\begin{figure*}
\centering
\includegraphics[width=\textwidth]{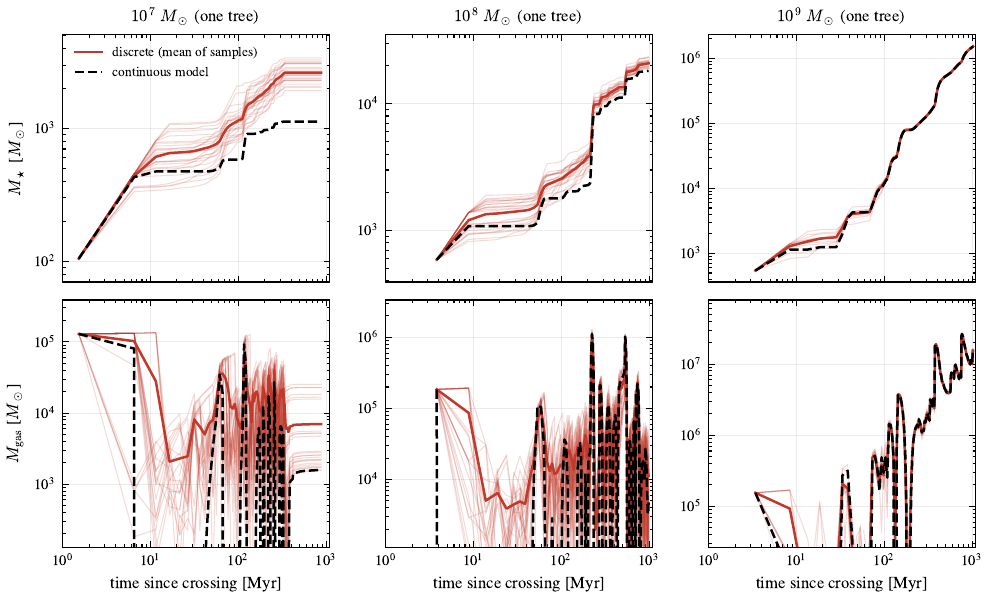}
\caption{Stellar mass (top) and gas mass (bottom) against time since star formation began, for one tree at each of three halo masses. Each tree is the one whose offset at $z=5$ is closest to the median. Thin red lines show individual discrete runs (25 at $10^7$ and $10^8\,\Msun$, 10 at $10^9\,\Msun$), the thick red line their mean, and the dashed line the IMF-averaged continuous model. Note that the gas mass drops to zero during several periods, and we choose not to plot the curves.}
\label{fig:evolution}
\end{figure*}

Figure~\ref{fig:evolution} shows the evolution of the stellar and gas masses in a representative tree at each of three halo masses. In the continuous model, gas removal begins about 4\,Myr after star formation starts, and the gas mass falls sharply at 8--9\,Myr. In the discrete runs, the gas remains at its initial value until the first supernova, whose arrival time varies between realisations. Their mean gas mass therefore declines later and more gradually. The discrete runs form more stars during this initial interval, and the resulting difference persists to $z=5$ at $10^7$ and $10^8\,\Msun$. At $10^9\,\Msun$, the stellar masses eventually converge.

The difference between the discrete and continuous models in mean stellar mass is substantial below about $10^8\,\Msun$ (as we will show in Figure~\ref{fig:offsetmass}). When energy that finds no gas is discarded, the mean stellar mass in the discrete runs exceeds that of the continuous model by 150, 135, 61 and 15 per cent at $5\times10^6$, $10^7$, $3\times10^7$ and $10^8\,\Msun$, respectively. At $10^9\,\Msun$, the difference is 0.1 per cent, and it is consistent with zero at $3\times10^9$--$3\times10^{10}\,\Msun$. The scatter between discrete runs of the same tree is much smaller than the difference between the two models (e.g. the scatter in stellar mass is 0.06\,dex at $10^7\,\Msun$; cf. Appendix~\ref{sec:appstats}).

\subsection{The onset and subsequent growth of the excess stellar mass}
\label{sec:when}

\begin{figure}
\centering
\includegraphics[width=\columnwidth,trim=0 0 239.61bp 14bp,clip]{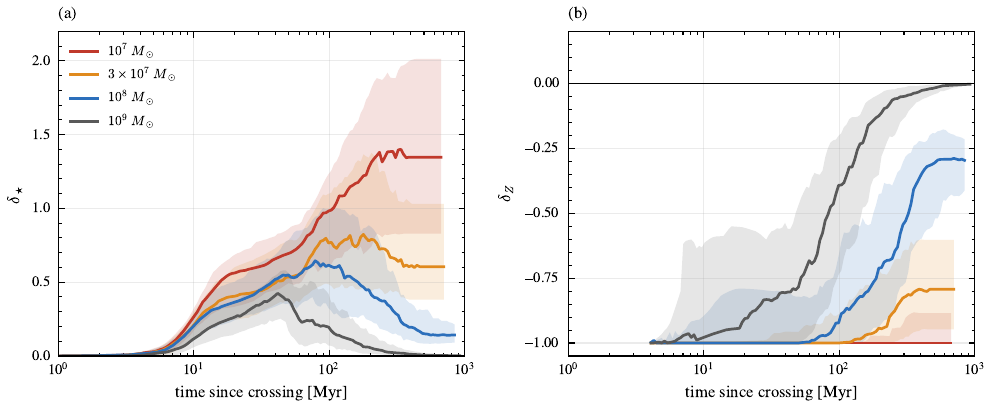}
\caption{Offset in stellar mass, $\delta_\star=\langle\Mstar\rangle/\Mstar^{\rm cont}-1$, against time since star formation began.
Curves show the median over trees, with the 16-84 per cent range.
}
\label{fig:offsettime}
\end{figure}

Figure~\ref{fig:offsettime} shows how the offset in stellar mass evolves with time since the halo's crossing of the molecular-cooling threshold ($7\times10^5\,\Msun$) for halos that reach masses between $10^7\Msun$ and  $10^9\Msun$ at $z$=5. The curves and shaded regions show the medians and 16--84 per cent ranges for each of the halo masses. The median curves of the lower masses (up to $10^8 \Msun$) trace each other to $\sim 50$ Myr; they all show a sharp rise to an offset of $\sim 25$ per cent at $\sim 10$ Myr and $\sim 50$ per cent at $\sim 20$ Myr, before rising gently to $\sim 60$ per cent at $\sim 50$ Myr. At this point they separate, and by $600$ Myr the $10^7\Msun$ median has risen to $\sim 135$ per cent; the $3 \times 10^7\Msun$ median has declined to $\sim 60$ per cent; and the $10^8\Msun$ median has dropped to $\sim 14$ per cent. This behaviour contrasts with the $10^9\Msun$ case, whose median curve departs from the lower masses at $\sim 5$ Myr, rises to a maximum offset of $\sim 40$ per cent at $\sim 45$ Myr, and then declines sharply to zero by $600$ Myr. If we evaluate the offset at the fixed redshift of $z=5$, we find values of 135, 15 and 0.15 per cent at $10^7$, $10^8$ and $10^9\,\Msun$ respectively. Trees reach $z$=5 at different times after crossing, so each curve ends before the first tree reaches $z$=5, at $\sim670$, 800, 840 and 920 Myr for $10^7$, $3\times10^7$, $10^8$ and $10^9\,\Msun$; the median at every plotted time is then over the same trees.

\begin{figure*}
\centering
\includegraphics[width=\textwidth]{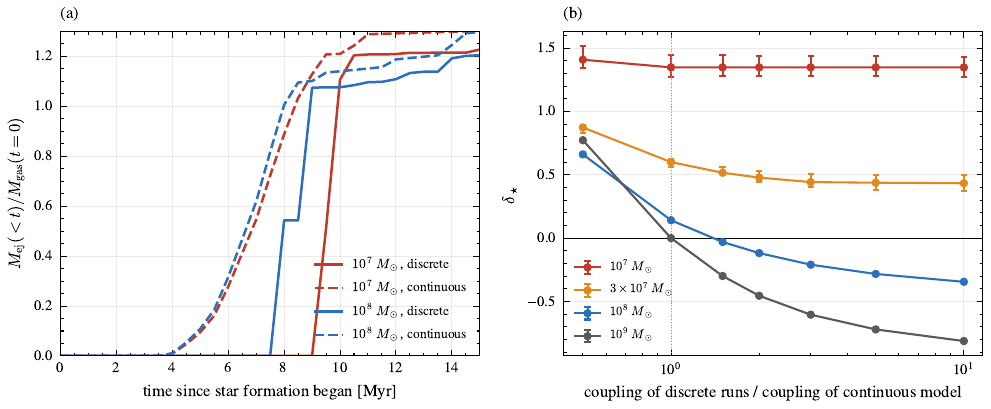}
\caption{(a) Gas removed by supernovae, in units of the gas present at the start, against time since star formation began: IMF-averaged continuous model (dashed) and discrete runs (solid), medians over 10 trees at $10^7$ (red) and $10^8\,\Msun$ (blue). (b) Offset in stellar mass, $\delta_\star$, at $z=5$ against the ratio of the coupling in the discrete runs to that in the continuous model (0.1), with the same energy per unit stellar mass formed. 
Error bars show the 16--84 per cent bootstrap interval of the median over trees.}
\label{fig:mechanism}
\end{figure*}

These trends can be understood by considering the size of the reservoir once the halo first starts forming stars and the first supernova occurs. At this point the gas mass is only 0.37--0.65 times the mass that one supernova can unbind. The continuous model begins removing gas after about 4\,Myr and empties the initial reservoir by 8--9\,Myr (Figure~\ref{fig:mechanism}a), when fewer than one supernova is expected. In contrast, the discrete model with the same expected number of events has had none with probability 0.52--0.69, and so no gas has been removed. On average, the first supernova occurs only after $1/\nu_{\rm SN} \simeq 85\,\Msun$ of stars have formed, where $\nu_{\rm SN}$ is the number of supernovae per solar mass of stars formed, and only after the progenitor's lifetime has elapsed; for this reason, in many realisations the first supernova occurs later than in the continuous model's expectation\footnote{A Poisson estimate of the first supernova, using the continuous model's cumulative event count, reproduces much of the early offset. It agrees with the measured offset at 20\,Myr to within 8 per cent at $10^7$, $3\times10^7$ and $10^8\,\Msun$.}. This means that star formation proceeds in the discrete model past the point at which it has been suppressed in the continuous model.

Interestingly, the duration of the delay to the first supernova does not set the final mass excess. If we consider the median tree at $10^7$--$10^8\,\Msun$, 76--83 per cent of the excess that forms comes after the first 50\,Myr. Instead, it reflects the gas mass that the discrete model does not remove from the reservoir. Both models follow the same tree, which governs the gas mass that accretes, and gas that the continuous model ejects but the discrete model does not ends up either as additional stars or as additional gas in the reservoir. The gas that supernovae can unbind per unit stellar mass formed is the same in the two models to within 6 per cent. However, between 100\,Myr and $z=5$, the discrete models remove only 29, 65 and 90 per cent as much gas per unit stellar mass formed as the continuous model at $10^7$, $3\times10^7$ and $10^8\,\Msun$, and the same amount at $10^9\,\Msun$. At $z=5$, 20--30 per cent of this difference has formed additional stars in the discrete model, with the rest as additional gas in the reservoir. We note that these additional stars form from gas that supernovae have not enriched, so their metallicity is lower than in the continuous model (see Appendix~\ref{sec:metals}).

\subsection{The origin of the excess stellar mass}
\label{sec:size}

The results so far suggest that the excess can be understood as a product of two related mechanisms: the finite gas mass available to an individual supernova event, and the delay until feedback occurs. Here we examine in more detail how these mechanisms contribute to the offset, whether they act independently, and whether there is any dependence on stochastic sampling of the supernovae or on the energy of individual events.

\medskip

\noindent\emph{Setting stellar lifetimes to zero tests} the sensitivity to the duration of the delay before the first supernova. At $z=5$ the offset rises from 135 and 61 per cent to 386 and 146 per cent at $10^7$ and $3\times10^7\,\Msun$, and is unchanged at $10^8\,\Msun$ (15 and 14 per cent). The continuous model removes gas immediately, whereas the discrete model must wait for its first supernova, which is expected only after sufficient stellar mass has formed (as shown in the previous subsection). This shows that the delay is set by the size of the event, not by the stellar lifetime.

\medskip

\noindent\emph{We assess the role of stochastic sampling} by replacing the Poisson distributed number of supernovae with its mean, the random delays with quantiles of the stellar-lifetime distribution, or both simultaneously. In every case we find that a large offset remains. At $10^8\,\Msun$ the offsets are 11.8, 18.2 and 18.9 per cent, compared with the baseline of 14.2 per cent, while at $10^7\,\Msun$ they are 126, 162 and 171 per cent, compared with 135 per cent. The standard errors of these medians are 0.7--0.9 and 6--9 per cent, respectively. This indicates that the offset is not caused by the particular stochastic realisation. With zero coupling, the stellar masses of the two models differ by less than 0.3 per cent at all three masses, which implies that the offset is driven by the removal of gas by supernovae.

\begin{figure*}
\centering
\includegraphics[width=\textwidth]{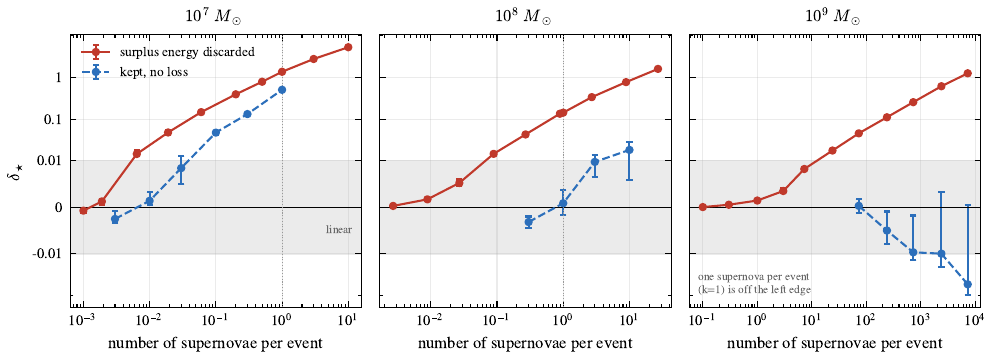}
\caption{Offset in stellar mass, $\delta_\star$, at $z=5$ against the energy carried by each event, in units of one supernova ($k$), with the same expected energy per unit stellar mass formed. The vertical axis is linear for $|\delta_\star|<0.01$ (shaded) and logarithmic beyond. Red points show runs in which energy that finds no gas is discarded; blue dashed points show runs in which it is retained without loss. Individual supernovae correspond to $k=1$ (dotted line), and the IMF-averaged limit to $k\rightarrow0$. Every point uses 100 trees. Error bars show the 16--84 per cent bootstrap interval of the median over trees, and every run uses independent random numbers in every tree.}
\label{fig:eventsize}
\end{figure*}

\medskip

Figure~\ref{fig:eventsize} examines how the energy carried by an individual event affects the offset at fixed energy per unit stellar mass formed, for halos of  $10^7$, $10^8$ and $10^9\,\Msun$ at $z$=5. We characterise the energy carried by an individual event using $k$, defined so that each event carries $k$ times the energy of one supernova and occurs at $1/k$ times the event rate needed to match the same total energy per unit stellar mass formed. Individual supernovae therefore correspond to $k=1$, and the IMF-averaged limit to $k\rightarrow0$; we use $k$ in this single sense throughout, including in Figure~\ref{fig:eventsize}.

If energy is lost because there is insufficient gas present in the reservoir during an event (red points), the offset decreases continuously towards zero as $k$ tends to zero. At $10^7\,\Msun$, it rises from 0.1 per cent at $k=0.002$ to 4.8, 39 and 135 per cent at $k=0.02$, 0.2 and 1. At $10^8\,\Msun$, it is 0.5 per cent at $k=0.03$, 14 per cent at $k=1$ and 76 per cent at $k=9$. The event size at which the offset falls below 1 per cent increases with halo mass: approximately 0.005, 0.06 and 10 supernovae at $10^7$, $10^8$ and $10^9\,\Msun$, respectively.

\medskip

These tests imply that the excess stellar mass we measure in the discrete model is not simply a consequence of stochastic sampling or finite stellar lifetimes, but is a signature of how and when supernova feedback injects energy into the reservoir in the respective models.

\subsection{The role of feedback coupling}
\label{sec:coupling}

If the excess is being driven by the manner in which supernova feedback injects energy into the reservoir, it is important to examine how the result is influenced by how efficiently the feedback couples to the reservoir. We investigate this in Figure~\ref{fig:mechanism}b, which shows how the offset at $z=5$ depends on the ratio of the coupling ($\epsfb$) in the discrete and continuous models, at fixed energy per unit stellar mass formed, for halos of $10^7$, $3\times10^7$, $10^8$ and $10^9\,\Msun$. Here the coupling ratio multiplies $\epsfb$, the fraction of the supernova energy available to unbind gas, in the discrete runs relative to the continuous model, while the supernova energy per unit stellar mass formed is unchanged. The ratio therefore measures the sensitivity of the offset to the relative coupling; it does not measure how efficiently individual remnants couple to the gas. Recall that a positive offset means the discrete model forms more stars than the baseline continuous model; below, some coupling ratios drive the offset negative, reversing this ranking.

The response depends strongly on halo mass. At $10^7\,\Msun$, an individual event unbinds substantially more gas than is present in the reservoir, so the first event empties it for every coupling ratio ($r=\epsfb/\epsfb^{\rm cont}$) tested, and the offset remains between 135 and 141 per cent for all values of $\epsfb$ tested. At $3\times10^7\,\Msun$, the offset falls from 60 to 43 per cent as the coupling ratio increases from 1 to 10, but remains substantial. At $10^8\,\Msun$, a ratio of about 1.4 eliminates the offset, while a ratio of 10 reverses its sign. At $10^9\,\Msun$, where star formation is regulated and the mass a single event can unbind, $m_1$, is small relative to the reservoir, the offset follows $1/r-1$ to within 9 per cent for coupling ratios $r=1.5$--5.

These differences reflect how the gas reservoir responds to changes in coupling. In the smallest halos, an individual event remains large relative to the available gas, so changing the relative coupling has little effect. In more massive halos, accretion can sustain a reservoir large enough that the relative coupling becomes increasingly important. Note, however, that our one-zone model cannot distinguish between events that couple more or less efficiently than the same energy distributed over smaller increments; this depends on the density and spatial structure of the gas and on radiative losses within the remnant, and realistically requires hydrodynamical simulations.

\begin{figure}
\centering
\includegraphics[width=\columnwidth]{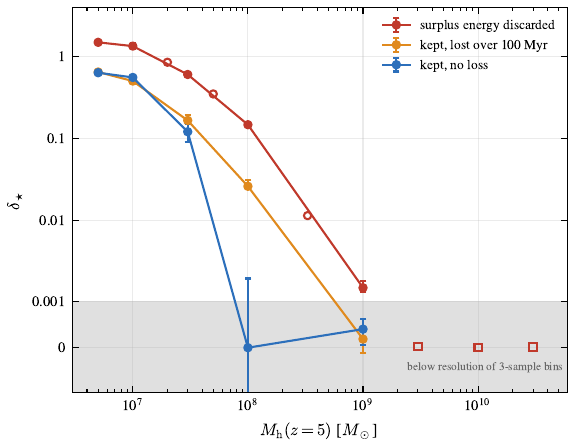}
\caption{Offset in stellar mass, $\delta_\star$, at $z=5$ against halo mass, when energy that finds no gas is discarded (red), retained for later inflow with a loss time of 100\,Myr (orange), or retained with no loss (blue). Filled points show the ensemble with different sets of random numbers in every tree (100 trees), with the 16--84 per cent bootstrap interval of the median over trees. Open red symbols show the fiducial model at the other masses, also with independent random numbers in every tree; the squares at $\geq3\times10^9\,\Msun$ use three discrete runs per tree, with 100 trees at $3\times10^9$ and $10^{10}\,\Msun$ and 54 at $3\times10^{10}\,\Msun$. The vertical axis is linear for $|\delta_\star|<0.001$ and logarithmic beyond, and the grey band indicates the resolution of those bins.}
\label{fig:offsetmass}
\end{figure}

We can gain some insight into the fate of energy that cannot couple to the gas during an event and its effect on the size of the offset by considering Figure~\ref{fig:offsetmass}. This considers three prescriptions for injected energy that is unused in an event: surplus energy is discarded (red); it is retained and couples to gas inflow with a loss time of 100\,Myr (orange); or it is retained with no loss (blue). At $10^7\,\Msun$, retaining unused injected energy reduces the offset from 135 per cent to 56 per cent when there is no energy loss, or 50 per cent for a loss time of 100\,Myr. At higher masses, the effect is smaller: the corresponding offsets are 12 and 17 per cent at $3\times10^7\,\Msun$, and 0.0 and 2.6 per cent at $10^8\,\Msun$. The two retention prescriptions differ by only 5 per cent at $10^7\,\Msun$.

The amplitude of the offset therefore depends not only on how much energy couples during an event, but also on the fate of energy that cannot couple to the available gas. Resolving this requires following the subsequent evolution of the gas and energy spatially, and realistically requires hydrodynamical simulations.

\subsection{Dependence on halo mass}
\label{sec:boundary}

The results so far show that the continuous model provides a reasonable approximation at higher halo masses. We now ask whether we can determine the halo mass at which the continuous approximation becomes adequate. We do this by identifying the halo mass at which the fractional offset in stellar mass at a given observation redshift falls below 10 per cent; this is our chosen operational criterion for the adequacy of the continuous approximation. The relevant physical quantity is the mass one supernova can unbind, $m_1=2\epsfb E_{\rm SN}/\Vesc^2$, relative to the gas in the reservoir. We consider this ratio at two stages: at the onset of star formation, where $m_1/\Mgas$ controls the initial offset, and in the regulated state that more massive halos reach, where the analogous diagnostic is $\Gamma_{\rm eq}$, introduced below.

At the onset of star formation, the median ratio $m_1/\Mgas$ is 1.5--2.7 across the halo mass range $10^7$--$10^9\,\Msun$ at $z=5$, assuming a halo mass threshold for cooling of $7\times10^5\,\Msun$. In this model, every halo begins with a reservoir that an individual event can empty, which is consistent with the similar excesses we observe in all halos at early times, regardless of their $z$=5 mass (cf. Figure~\ref{fig:offsettime}). The ratio $m_1/\Mgas$ depends on the halo mass threshold for star formation; raising the threshold increases the initial reservoir and reduces the offset smoothly. At $10^7\,\Msun$ it falls from 135 to 68 and 24 per cent as the threshold rises from $0.7$ to $1.4$ and $2.8\times10^6\,\Msun$ (Appendix~\ref{sec:appregime}). Offsets of 9--28 per cent persist even when the onset ratio falls to 0.1--0.3, so the difference continues to develop after star formation begins.

In more massive halos, the reservoir is determined by the balance between accretion, star formation, and feedback. In this regulated regime, we can estimate the equilibrium gas mass relative to $m_1$ from the accretion history. Neglecting stellar lifetimes, accretion balances star formation and supernova-driven gas removal, giving
\begin{equation}
\Gamma_{\rm eq}=\frac{M_{\rm g,eq}}{m_1}
=\frac{\dot M_{\rm acc}\,\tdyn}
{\epsff\,(1+\nu_{\rm SN}m_1)\,m_1}.
\label{eq:gammaeq}
\end{equation}
At $10^8$ and $10^9\,\Msun$, this estimate reproduces the median gas mass of the continuous model, in units of $m_1$, to within 0.01 and 0.13\,dex, giving $\Gamma_{\rm eq}=0.39$ and 460, respectively. At $10^7$ and $3\times10^7\,\Msun$, it overestimates the gas mass by 2.5 and 1.8\,dex. These halos do not reach the regulated state because individual events repeatedly empty the reservoir. 

These two measures describe different stages of the evolution. $m_1/\Mgas$ characterises the fragile initial reservoir and $\Gamma_{\rm eq}$ describes the reservoir that can be sustained in more massive, regulated halos. This distinction helps to explain the mass dependence of the final offset. At $z=5$, it falls from 62 per cent at $3\times10^7\,\Msun$ to 15 per cent at $10^8\,\Msun$, and becomes negligible at $10^9\,\Msun$. In the regulated regime, the offset is 15 per cent at $\Gamma_{\rm eq}\approx0.4$ and becomes negligible when $\Gamma_{\rm eq}\gg1$. The value of $\Gamma_{\rm eq}$ at the 10 per cent crossing is not universal, however, ranging from 0.13 to 1.2 across the redshift and coupling combinations with $\epsfb\geq0.05$. At $\epsfb=0.01$ only $z_{\rm meas}=5$ reaches 10 per cent, where the offset lies between 8 and 18 per cent over $1.6\times10^6$--$1.1\times10^7\,\Msun$, so the crossing is poorly defined.

\begin{figure*}[t]
\centering
\includegraphics[width=\textwidth]{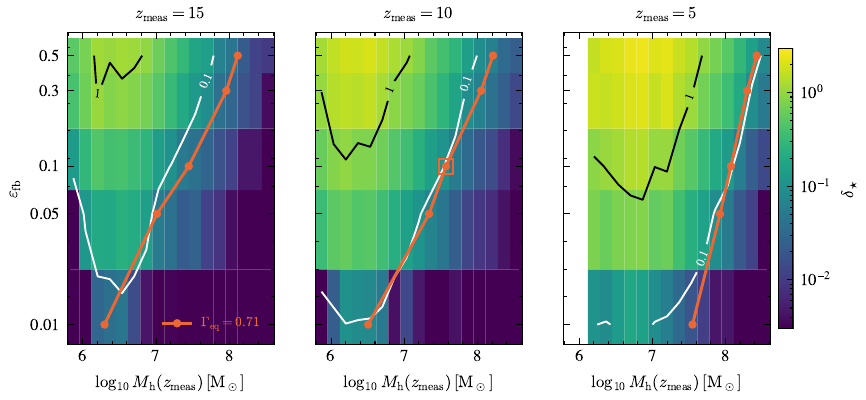}
\caption{Median offset in stellar mass, $\delta_\star=\langle\Mstar\rangle/\Mstar^{\rm cont}-1$, from the IMF-averaged continuous model (colour, logarithmic), for halos at the molecular threshold, against halo mass at measurement redshift $z_{\rm meas}$ and coupling $\epsfb$. White and black contours mark $\delta_\star=0.1$ and 1. The orange line shows the halo mass at which $\Gamma_{\rm eq}=0.71$, calibrated at the 10 per cent crossing for $z_{\rm meas}=10$ and $\epsfb=0.1$ (square). The estimate is calculated at each redshift and coupling from the accretion history, escape velocity and coupling alone, using the first 16 trees of each cell. At $\epsfb=0.01$ the offset remains below 10 per cent at $z_{\rm meas}=15$ and 10, so there is no boundary to compare there. The grid uses the Chabrier IMF adopted throughout.}
\label{fig:regime}
\end{figure*}

Figure~\ref{fig:regime} shows the offset (colour) against halo mass at the measurement redshift, for $z_{\rm meas}=15$, 10 and 5 (panels) and couplings $\epsfb=0.01$--0.5. The white contour marks the 10 per cent criterion and identifies the halo mass at which the continuous approximation becomes adequate. We compare this with an estimate based on $\Gamma_{\rm eq}$, calculated from the accretion history, escape velocity, and coupling. The orange line shows the mass at which $\Gamma_{\rm eq}=0.71$, calibrated to the 10 per cent crossing at $z_{\rm meas}=10$ and $\epsfb=0.1$ (square). This provides a simple estimate of how the mass at which the continuous approximation becomes adequate varies with redshift and coupling.

\subsection{Recovery of the continuum limit}
\label{sec:continuum}

We find that the continuous model can be recovered in two limits. For a fixed event size, the difference in final stellar mass becomes negligible as halo mass increases; it is 0.15 per cent at $10^9\,\Msun$ and consistent with zero at higher masses (Figure~\ref{fig:offsetmass}). For a fixed halo mass, the difference tends to zero as the energy carried by each event decreases (Figure~\ref{fig:eventsize}). 

% ============================================================
\section{Discussion}
\label{sec:discussion}
% ============================================================

\subsection{The physical origin of the excess stellar mass}

A supernova is dynamically important when the gas mass it can unbind, $m_1$, is comparable to or larger than the available reservoir. In our model, this condition is met at the onset of star formation in every halo between $10^7$ and $10^9\,\Msun$, and we find that the initial reservoir is only 0.37--0.65 of $m_1$. This effect is most pronounced in the smallest halos; the median gas mass in the $10^7\,\Msun$ continuous model is just 0.5 per cent of $m_1$.

Our models for discrete and continuous feedback differ in two fundamental respects. First, they treat gas removal from the finite reservoir differently. An individual event cannot remove more gas than is present: when $m_1$ exceeds the mass in the reservoir, the event removes all the gas and any remaining energy has nothing to act on. In the continuous prescription, by contrast, the same energy is distributed over time, allowing gas to be removed in smaller increments while the reservoir remains. This means that gas removal is not proportional to the feedback energy injected when the available reservoir is finite. 

Second, the prescriptions differ in when feedback first acts. In the continuous prescription, feedback begins removing gas about 4\,Myr after star formation begins, whereas the first discrete event occurs after 8--10\,Myr, once about $85\,\Msun$ of stars have formed. This means that star formation continues for a more extended period under discrete feedback. The offset persists when the number and timing of supernovae are fixed, and tends to zero as the energy per event does. The continuous prescription therefore does not generally reproduce the outcome of individual events: averaging the energy input before applying the gas-removal limit gives a different result from applying the limit to individual events and then averaging the outcomes.

For a fixed reservoir, an event capable of ejecting $m$ removes $\min(M_{\rm g},m)$. This response saturates at $M_{\rm g}$, and so large events cannot remove gas in proportion to their energy; in other words,
\[
\langle\min(M_{\rm g},m)\rangle
\leq \min(M_{\rm g},\langle m\rangle).
\]
In this way, averaging the event size before applying the gas-removal limit gives at least as much gas removal as averaging the outcomes of individual events. The inequality illustrates the origin of the offset but does not determine the evolution; this also depends on gas accretion and star formation between events, and the delay before the first supernova takes effect. Equality is approached when individual events are small compared with the reservoir, as in the continuous limit as $k$ tends to zero, and in the most massive halos. This argument assumes that surplus energy is lost; the retained-energy and coupling variants change the amplitude and potentially the sign of the offset (\S\ref{sec:coupling}).

We tested the small-event limit using a continuous model that releases the energy from each star-formation episode in $N_{\rm bin}=200$ increments, at the lifetimes of the corresponding progenitor masses, and applied this gas-removal limit to each increment. This retains the stellar lifetime delay while reducing the energy injected per event. The offset tends to zero as the energy per event decreases (\S\ref{sec:size}), as we would expect for the small-event limit. We also compared the discrete runs with an alternative continuous model (Appendix~\ref{sec:appregime}). A continuous model that releases the energy when the stars form, with no lifetime delay, forms 0.28, 0.41 and 0.66 times the stellar mass of the baseline continuous model at $10^7$, $3\times10^7$ and $10^8\,\Msun$, respectively. The corresponding offsets rise from 135, 61 and 15 per cent to about 700, 290 and 70 per cent. The baseline continuous model is the conservative choice in this comparison.

\subsection{Dependence on halo mass}

Once star formation begins, we expect the evolution of the reservoir to depend on halo mass: gas accretion tracks the halo mass growth rate, and the efficiency of gas removal depends on the escape velocity of the halo at that time. In the lowest-mass halos, individual supernova events can repeatedly empty the reservoir before accretion replenishes it, preventing the system from reaching a regulated state. The initial ratio $m_1/\Mgas$ provides a guide to how strongly the reservoir is affected by the first event, but does not determine the eventual difference in stellar mass between the discrete and continuous models. This difference depends on the subsequent competition between accretion and gas removal, as well as the delay before the first event.

Star formation begins when the halo reaches a prescribed mass threshold, which sets the initial gas mass available when the first supernova occurs, and hence the ratio $m_1/\Mgas$. If we increase the halo mass threshold for when stars can form, this means that more gas will accumulate in the halo before feedback occurs, and a smaller fraction  of the reservoir will be ejected by the first supernova. This leads to a reduced stellar-mass offset between the discrete and continuous models. In our fiducial model, we assume that the halo mass threshold is based on the requirement for molecular cooling, whose value we take from recent simulations \citep[cf.][]{Lenoble2024}. For a halo that reaches $10^7 \Msun$ at $z=5$, increasing the halo mass threshold from the fiducial value of $0.7 \times 10^6\Msun$ by factors of 2 and 4  (i.e. $1.4$ and $2.8\times10^6\,\Msun$) reduces the offset from 135 to 68 and 24 per cent, respectively (Appendix~\ref{sec:appregime}). The halo mass at which the discrete and continuous models converge will be sensitive to its choice (as we show when considering an atomic cooling mass threshold as an alternative).

Once halos reach a regulated state, the equilibrium gas mass provides a useful estimate of the halo mass above which the continuous approximation reproduces the discrete results to within 10 per cent (\S\ref{sec:boundary}). We introduced the quantity $\Gamma_{\rm eq}=M_{\rm g,eq}/m_1$ (Equation~\ref{eq:gammaeq}), which compares the equilibrium gas mass supported by accretion with the mass that can be unbound by a single supernova. At $z=5$ and $\epsfb=0.1$, $\Gamma_{\rm eq}$ is 0.4 at $10^8\,\Msun$ and exceeds 400 at $10^9\,\Msun$, as the equilibrium reservoir becomes progressively larger relative to the mass unbound by an individual event. We find that the fractional stellar-mass offset between the discrete and continuous models falls below 10 per cent at $\Gamma_{\rm eq}=0.71$ in our calibration case. Adopting this value predicts the halo mass at which the offset falls below 10 per cent to within a factor of two in 9 of the 12 redshift and coupling cases tested. The exceptions are $z_{\rm meas}=15$ at $\epsfb=0.3$ and 0.5 and $z_{\rm meas}=5$ at $\epsfb=0.01$, where it overestimates the mass by factors of 2.3, 2.4 and 3.8. This suggests that, within this model, $\Gamma_{\rm eq}$ is a useful first estimate of where the continuous approximation becomes reasonable, although it overestimates this mass at $z_{\rm meas}=15$ for $\epsfb\geq0.3$. We stress that the 10 per cent criterion is a convenience adopted for this paper, not a physically significant threshold; a different choice would shift the inferred halo mass smoothly, as Figure~\ref{fig:regime} and Appendix~\ref{sec:appregime} show.

\subsection{Dependence on feedback coupling}

The magnitude and sign of the stellar-mass offset between discrete and continuous feedback depend on how supernova energy couples to the gas. In our model, the coupling efficiency is identical in both treatments, and energy in excess of that required to remove the available gas is lost. In a resolved interstellar medium, however, individual events may couple more efficiently than smaller energy deposits if they heat gas before radiative losses can remove the energy. Consistent with this possibility, clustered supernovae can deliver substantially more momentum per event than isolated explosions \citep{Gentry2017} and drive stronger winds \citep{Fielding2018}. In our coupling-ratio tests, increasing the coupling of discrete events by a factor of about 1.4 eliminates the stellar-mass offset at $10^8\,\Msun$. At $3\times10^7\,\Msun$, increasing the ratio to 10 reduces the offset but does not reverse it, while at $10^7\,\Msun$ none of the ratios tested reverses the result.

The fate of energy in excess of that required to remove the available gas also matters. Retaining this energy for coupling to subsequently accreted gas substantially reduces the offset in the smallest halos, but does not reverse its sign in the variants tested. Replacing energy-driven with momentum-driven gas removal reduces the offset by about a factor of three at $10^7$--$10^8\,\Msun$, but leaves a 39 per cent offset at $10^7\,\Msun$. Thus, although the magnitude of the effect depends on the feedback prescription, the excess stellar mass produced by discrete feedback persists in the smallest halos across the alternatives we have examined.

\subsection{The survival of gas until the first supernova}
Our results apply to halos that retain a substantial gas reservoir until the first supernova explodes. This is not a secondary caveat: the entire effect we study presupposes that such a reservoir survives, so whether it does is one of the two dominant physical uncertainties in our argument, the other being the relative coupling of discrete and continuous feedback to the gas (\S\ref{sec:coupling}). Radiation from the first stars could remove or redistribute this gas before supernova feedback begins, but our model does not follow this process explicitly. We include suppression of gas accretion by the UV background after reionisation and represent the effect of a Lyman–Werner background through our fixed star formation threshold, but we do not model photoionisation by the first stars.

The escape velocities of halos when star formation can proceed are 5--9\,km/s, and below about 11\,km/s in halos with offsets above 8 per cent. These are comparable to or below the sound speed of fully photoionised gas, 11.7 km/s. For this reason, photoheating could substantially alter the reservoir before the first supernova. \citet{Kitayama2004} find that ionisation fronts escape halos below about $10^6\,\Msun$ within a few $10^5$\,yr for a single massive star, whereas they remain inside the virial radius in halos of $10^7\,\Msun$ and above. In the fiducial case, our halos have masses of $7\times10^5\,\Msun$ at the onset of star formation, but whether the small stellar populations that form can produce sufficient radiation to remove the gas is beyond the scope of our model. The radiation field would also change the density and temperature of the remaining gas, potentially altering the coupling of subsequent supernovae.

We test the sensitivity to early gas removal with a simplified prescription that removes gas in both treatments at a rate proportional to the mass of stars younger than 4 Myr. For rates up to 30\,$\Msun$ of gas per $\Msun$ of young stars per Myr, the offsets at $10^7$--$10^8\,\Msun$ change by at most 10 per cent. At a rate that is 10 times higher, they fall to 34, 31 and 16 per cent at $10^7$, $3\times10^7$ and $10^8\,\Msun$, respectively, although the continuous model then forms less than half its original stellar mass. Note that this simple prescription does not model photoionisation or its effect on supernova coupling, but is designed to assess whether early gas removal can reduce the offset.

\subsection{Comparison with previous studies}

The difference between discrete and continuous feedback depends on how supernovae are sampled; how their energy is injected and coupled to the gas; and how fluctuations in star formation and outflow rates are modelled. These distinctions matter when comparing our results with simulations that sample the stellar initial mass function (IMF) or resolve individual supernovae.

Previous studies have examined different aspects of feedback variability: IMF sampling and supernova timing; the injection and coupling of supernova energy; and fluctuations in star formation and outflow rates. We note that they do not directly test the response of a finite gas reservoir to individual supernova events that we investigate here.

\medskip

\noindent\emph{IMF sampling and supernova clustering.} \citet{Applebaum2020} find that a stochastically sampled IMF typically gives lower stellar masses in halos below about $10^{8.5}\,\Msun$, sometimes by more than an order of magnitude, because feedback becomes burstier and quenches star formation earlier. This is opposite to the effect in our model. More efficient coupling of individual events could account for the difference, but establishing this requires resolving the gas response to the supernova remnant. \citet{JeonGo2026}, by contrast, find similar final stellar masses for burst, stochastic, and individual sampling in three halos of $10^8\,\Msun$ at $z=0$, although individual sampling produces stellar metallicities higher by 0.2--0.5 dex. Their burst model releases the energy of an entire stellar population at once and is not equivalent to our IMF-averaged continuous model. Indeed, their Table~1 shows that the burst model produces the highest or equal-highest stellar mass in all three halos.

\medskip

\noindent\emph{Energy injection and coupling.} \citet{Su2018} find, in dwarf galaxies in halos of $10^9$--$10^{10}\,\Msun$, that the discreteness of individual supernovae is important for regulating stellar masses and driving galactic winds, and that once supernovae are discrete, IMF sampling has weak effects on galaxy-scale properties. \citet{Smith2021IMF} find no difference between sampled and averaged rates of discrete supernovae when supernovae are the only feedback channel. These results illustrate that the consequences of sampling the supernova population depend on how energy injection is represented, but they do not establish how individual events couple to a finite gas reservoir.

\medskip

\noindent\emph{Stochastic variations in star formation and outflows.} The fluctuations imposed by \citet{PanKravtsov2023} and \citet{KravtsovBelokurov2024} produce variations of 1--2 magnitudes in UV luminosity at $z=10$--16 in halos of $\geq10^9\,\Msun$, where our offset is below 1 per cent. However, their UV luminosity fluctuations are not directly comparable to our stellar mass offsets. Their prescriptions vary the outflow rate with the star formation history, whereas our model follows the response of a finite gas reservoir to individual supernova events. In our model the difference persists when the number and timing of supernovae are fixed, showing that IMF sampling is not required for the effect.

\subsection{Implications for galaxy formation models and simulations}

For the final stellar mass at $z=5$, the continuous approximation is adequate at equal coupling in halos of $10^9\,\Msun$ and above. This agreement does not extend to their early evolution; the stellar masses accumulated during the first 50\,Myr differ at every halo mass, and the metal content can also differ substantially at early times in halos whose stellar masses subsequently converge (Appendix~\ref{sec:metals}). Between $5\times10^6$ and $10^8\,\Msun$, the continuous treatment underestimates the final stellar mass by 15--150 per cent. A constant feedback coupling cannot compensate for this difference. Matching the mean stellar masses of the discrete runs requires the coupling in the continuous model to fall from 0.1 at $10^9\,\Msun$ to at most 0.012 at $10^7\,\Msun$ and 0.006 at $5\times10^6\,\Msun$. This mass dependence is steeper than the one-third power expected for a coupling proportional to escape velocity. Neither a constant coupling nor a momentum-like scaling can recover the discrete result.

Our results have implications for both semi-analytical models and cosmological simulations of galaxy formation. For semi-analytical models, they suggest several quantities that should be tracked when comparing discrete and continuous feedback: the gas mass available when the first supernova occurs, the delay between the onset of star formation and the first event, and the ratio of the mass an individual event can unbind to the available gas reservoir. These quantities indicate when feedback becomes limited by the finite gas supply, and when an energy-injection rate averaged over the IMF may overestimate gas removal. The treatment of energy that cannot be coupled to the available gas is also important.

The issue is also relevant to cosmological hydrodynamical simulations, some of which are beginning to model stellar populations and feedback on a star-by-star basis. The Aeos simulations \citep{Brauer2025}, for example, represent individual stars and follow their radiative and mechanical feedback, in contrast to comparison simulations that represent stellar populations with more massive particles; comparing star-by-star and IMF-averaged particle implementations directly, \citet{BrauerIMF2025} find that star-by-star modelling produces systematically less bursty feedback, consistent with the sense of the offset we report here. \citet{Gutcke2021} present a related implementation, \textsc{lyra}, in which individually resolved supernovae with variable energy and self-consistent spatio-temporal clustering drive significant galactic winds; their model does not isolate this from an averaged treatment directly, but it demonstrates the same individually resolved, clustered injection that our discrete model represents in a single zone. A complementary, resolution-based perspective comes from \citet{Smith2026}, whose star-by-star model finds that supernova clustering produces burstier star formation and stronger winds once the mass resolution exceeds about $100\,\Msun$ per star particle; this mass-resolution threshold for discreteness effects is conceptually analogous to the halo-mass threshold we identify here, although the two are not directly comparable. \citet{Zhang2025} add a further, complementary resolution result: in isolated dwarf simulations, the local- and large-scale channels through which supernovae suppress star formation and drive outflows are only distinguishable at particle masses finer than about $10^2\,\Msun$; at coarser resolution the two channels become mixed together and indistinguishable. 

Such calculations offer the opportunity to examine how the treatment of individual feedback events affects the evolution of low-mass galaxies. However, the consequences will depend on how star formation is resolved, how feedback is coupled to the gas, and how the subsequent evolution of supernova remnants is captured. Our one-zone results do not prescribe a mass-resolution requirement for these simulations. Rather, they identify a specific issue to test: whether the gas response to an individual supernova differs from the response to the same energy distributed over smaller increments.

% ============================================================
\section{Conclusions}
\label{sec:conclusions}
% ============================================================

Galaxy formation models commonly represent supernova feedback as a continuous rate of gas removal, averaged over the stellar initial mass function (IMF). This assumes that feedback can be averaged over individual events without changing the resulting stellar mass. We have tested this assumption using a one-zone model that follows gas accretion along Monte Carlo merger-tree histories, comparing discrete supernovae with a continuous model that injects the same energy per unit stellar mass formed. By limiting the gas that an individual event can eject to the reservoir present at the time of the event, our model isolates the effects of finite reservoirs and discrete energy injection. We compare our results with the baseline continuous model (\S\ref{sec:model}); releasing the energy when the stars form increases the offset (\S\ref{sec:discussion}).

We find that the continuous approximation can fail substantially in low-mass halos. At $10^7\,\Msun$, the discrete model forms 135 per cent more stars by $z=5$ than the continuous model; at $10^8\,\Msun$ the excess is 14 per cent, while at $10^9\,\Msun$ the two agree to 0.1 per cent. The difference arises because an individual supernova can inject more energy than the reservoir can absorb. The offset is associated with the nonlinear response of the reservoir, through the gas-mass limit and the cessation of star formation. In the fiducial model the excess energy is lost; retaining it for later inflow reduces the offset at $10^7\,\Msun$ from 135 to 50--56 per cent but does not change its sign. Continuous feedback distributes the same energy across smaller increments and can couple it to gas that remains in the reservoir. The delay before the first discrete event sets when the two models begin to diverge, but the finite gas reservoir sets how large that divergence becomes. Fixing the number and timing of supernovae leaves the offset largely unchanged, so the difference is not an IMF-sampling effect. It is a systematic effect, and it vanishes as the energy per event tends to zero.

The difference depends on the reservoir relative to the energy of an individual event, not on halo mass alone. It declines smoothly as the reservoir grows relative to $m_1$, with no sharp transition when the two masses become equal. The final excess stellar mass is largely assembled after the first 50\,Myr, from gas retained by the discrete runs. If we calibrate to agree at $10^9\,\Msun$, the continuous model underestimates stellar masses by 0.37\,dex at $10^7\,\Msun$ and 0.06\,dex at $10^8\,\Msun$. The scatter between discrete realisations is much smaller than these shifts, indicating that the principal consequence is a change in the mean stellar mass rather than an increase in its scatter.

Whether this difference matters for real dwarf galaxies depends on two unresolved aspects of the physics. First, halos must retain gas until their first supernova, despite photoionisation by the ultraviolet background and their own massive stars. Second, the relative coupling of a single remnant and the same energy delivered in smaller increments must be established in a realistic, multiphase interstellar medium. A change of about 1.4 in their relative coupling is sufficient to remove the offset at $10^8\,\Msun$, while a resolved simulation \citep{Applebaum2020} has found the opposite sign.

These results motivate simulations that resolve individual supernova remnants and the reservoirs in which they expand. Comparing otherwise identical calculations with energy injected in a single event or in smaller increments would directly measure the relative coupling. Connecting the effect to present-day dwarf galaxies will require following the calculations beyond $z=5$, with pre-enrichment, reionisation, infall and stripping. More generally, our results show that equal feedback energy per unit stellar mass does not necessarily imply equal galaxy formation. The discreteness of energy injection and the finite reservoir available to absorb it can matter as much as the amount of energy released.

% ============================================================
\appendix
% ============================================================

\section{Dependence on model parameters}
\label{sec:appregime}

We test the sensitivity of the offset to the feedback coupling, the star formation threshold, the treatment of energy that cannot couple to the available gas, and the prescription for gas removal. Figure~\ref{fig:regime} shows the dependence on measurement redshift and feedback coupling, using the Chabrier IMF. At $\epsfb=0.1$, the largest offsets are 46, 95 and 138 per cent at $z_{\rm meas}=15$, 10 and 5, respectively, occurring at halo masses of $1.6\times10^6$, $1.6\times10^6$ and $7.4\times10^6\,\Msun$. At $10^8\,\Msun$, the offset is 1.2 and 1.9 per cent at $z_{\rm meas}=15$ and 10, but remains at 13 per cent at $z_{\rm meas}=5$. At $3.4\times10^8\,\Msun$, it is below 1 per cent at $z_{\rm meas}=15$ and 10, and 1.4 per cent at $z_{\rm meas}=5$. Increasing the coupling increases the offset. Above $\epsfb\simeq0.01$, it scales approximately as $\epsfb^{0.4\text{--}1.2}$, with shallower slopes at lower halo masses and later redshifts. At $\epsfb=0.01$ the offset is at most 4.5, 9.6 and 18 per cent at $z_{\rm meas}=15$, 10 and 5.

\medskip

\noindent\textbf{Star formation threshold.} This changes the gas mass available when star formation begins. At $10^7\,\Msun$, increasing the threshold from $0.7$ to $1.4$ and $2.8\times10^6\,\Msun$ reduces the offset from 135 to 68 and 24 per cent, respectively. At $10^8\,\Msun$, increasing it from $0.7$ to $1.4$, $2.8$ and $5.6\times10^6\,\Msun$ reduces the offset from 15 to 13, 9 and 8 per cent. At $3\times10^7\,\Msun$, the offset falls from 60 to 12 per cent as the threshold increases from $0.7$ to $5.6\times10^6\,\Msun$. The halo mass at which the offset falls below 10 per cent therefore depends on the initial gas reservoir as well as on halo mass.

\medskip

\noindent\textbf{Retention of feedback energy.} The treatment of energy that cannot couple to the available gas affects the offset. At $10^7\,\Msun$, it falls from 135 per cent when surplus energy is discarded to 56 per cent when it is retained without loss, and 50 per cent when it is retained with a loss time of 100\,Myr. The corresponding values are 60, 12 and 17 per cent at $3\times10^7\,\Msun$, and 14, 0.0 and 3 per cent at $10^8\,\Msun$. The two retention prescriptions differ by only 5 per cent at $10^7\,\Msun$, and their ordering is not resolved. Thus, retaining energy for subsequent inflow reduces the offset, particularly at lower halo masses, although the precise retention prescription has little resolved effect.

\medskip

\noindent\textbf{Energy- versus momentum-driving.} We tested a momentum-driven prescription for gas removal following \citet{MenonBaluPower2026}. The resulting offset is about 0.3 times the energy-driven value at halo masses between $5\times10^6$ and $10^8\,\Msun$, giving offsets of 39 per cent at $10^7\,\Msun$ and 5.0 per cent at $10^8\,\Msun$. The halo mass at which the offset falls below 10 per cent shifts from $1.2\times10^8$ to $5\times10^7\,\Msun$ with this prescription, and to $1.5\times10^8\,\Msun$ when the momentum is increased by a factor of three.

\medskip

\noindent\textbf{Continuous feedback model.} We tested the dependence of the offset on the continuous feedback prescription, keeping the discrete runs unchanged. We use 100 trees and 25 discrete realisations per tree. The baseline continuous model gives offsets of 135, 61 and 15 per cent at $10^7$, $3\times10^7$ and $10^8\,\Msun$, respectively. Releasing the energy when the stars form, with no stellar-lifetime delay, gives stellar masses of 0.28, 0.41 and 0.66 times those of the baseline continuous model, and offsets of about 700, 290 and 70 per cent. This alternative removes gas earlier, which reduces the stellar mass formed in the continuous model and increases the offset.

% ============================================================
\section{Numerical accuracy}
\label{sec:appnum}
% ============================================================

The measured offset could be affected by three numerical choices: the resolution of the merger trees; the gas supplied to the two treatments; and the time integration. We test the sensitivity of the offset to the tree resolution, the time step, the discretisation of the continuous model and the stellar lifetimes, and Table~\ref{tab:numerical} lists the effect of each test on the stellar mass of the continuous model and on the offset. The offset is insensitive to all of these choices. The gas supply and the order of star formation and supernovae within a substep are fixed by construction, as described below.

\medskip

\noindent\textbf{Merger-tree resolution.}
The star formation threshold of $7\times10^5\,\Msun$ is resolved by 47 tree resolution elements. Refining the trees by a factor of four leaves the ratio $m_1/\Mgas$ at the onset of star formation unchanged. With a power-spectrum cutoff that represents both $\sigma(M)$ and its slope accurately, this refinement changes the discrete--continuous offset by $-7\pm9$ per cent at $10^7\,\Msun$ and by $-0.9\pm0.7$ per cent at $10^8\,\Msun$. Changing the cutoff alone at the standard tree resolution changes the offsets by $+2\pm10$ and $+0.2\pm0.8$ per cent. These shifts are comparable to their uncertainties and small compared with the variation between assembly histories, so the offsets are not sensitive to the tree resolution or the cutoff.

\medskip

\noindent\textbf{Gas supply.}
The halo mass in the merger trees increases in steps at mergers, and substeps are split at supernova times, which differ between the two treatments. To ensure that both treatments receive the same gas, we compute the supply from a cumulative table of the suppressed accretion, built once for each tree and shared by the two treatments, with the suppression evaluated at the mass at the mid-point of each table interval. The two treatments then receive the same total gas to better than $10^{-7}$ of the total. The convention for evaluating the suppression changes the total itself by 0.1--0.5 per cent, but it does so in the same way for both treatments and so does not contribute to the offset.

\medskip

\noindent\textbf{Time integration.}
Star formation and the supernovae it produces must occur in the correct order within a substep. We solve the star formation equation exactly in each substep; measure supernova times from the mean formation time of the stars formed; and cap the substep at 1\,Myr, shorter than the minimum stellar lifetime of 3.5\,Myr. Reducing the cap to 0.25\,Myr changes the stellar mass of the continuous model by 0.12 per cent at $10^7\,\Msun$ and 0.01 per cent at $10^8\,\Msun$, and the offset by $-3$ and $-1$ per cent, which is less than one standard error in each case. The stellar mass of the continuous model also changes by less than 0.1 per cent between $N_{\rm bin}=200$ and 800, and the offset by $-0.2$ and $+0.1$ per cent. The offsets are therefore converged in the time step and in the discretisation of the continuous model.
\medskip

\noindent\textbf{Stellar lifetimes.}
Using the lifetimes at $Z=0.014$ \citep{Ekstrom2012} in place of the fiducial lifetimes at $Z=0.002$ \citep{Georgy2013} changes the stellar mass of the continuous model by 3.2 per cent at $10^7\,\Msun$ and 0.9 per cent at $10^8\,\Msun$, but the offset by only $-1.8$ and $-0.6$ per cent, well within one standard error. The offset between the discrete and continuous models changes by less than one standard error, despite the change in the continuous-model stellar mass.

\begin{table*}
\centering
\caption{Effect of the numerical tests of Appendix~\ref{sec:appnum}, relative to the adopted calculation, on the stellar mass of the continuous model and on the offset $\delta_\star$ at $10^7$ and $10^8\,\Msun$. Each row changes one parameter, and the adopted values are given in the text. Standard errors are shown where available; a dash means the quantity was not tabulated.}
\label{tab:numerical}
\begin{tabular}{lcccc}
\hline
 & \multicolumn{2}{c}{Stellar mass of continuous model (per cent)} & \multicolumn{2}{c}{Offset (per cent)} \\
Test & $10^7\,\Msun$ & $10^8\,\Msun$ & $10^7\,\Msun$ & $10^8\,\Msun$ \\
\hline
\multicolumn{5}{l}{\emph{Merger-tree resolution}} \\
Trees refined $\times4$ & --- & --- & $-7\pm9$ & $-0.9\pm0.7$ \\
Power-spectrum cutoff alone & --- & --- & $+2\pm10$ & $+0.2\pm0.8$ \\
\multicolumn{5}{l}{\emph{Time integration}} \\
Substep cap 0.25\,Myr & $-0.12$ & $-0.01$ & $-3.2$ & $-1.1$ \\
$N_{\rm bin}=800$ IMF bins & $+0.01$ & $-0.01$ & $-0.2$ & $+0.1$ \\
\multicolumn{5}{l}{\emph{Stellar lifetimes}} \\
Lifetimes at $Z=0.014$ & $+3.2$ & $+0.9$ & $-1.8$ & $-0.6$ \\
\hline
\end{tabular}
\end{table*}

% ============================================================
\section{Merger and supernova histories}
\label{sec:appstats}
% ============================================================

We find that the results vary between halo merger trees because their assembly and gas-accretion histories differ, and between discrete realisations of a given tree because the number and timing of supernovae differ. Each discrete run uses its own set of random numbers, fixed by its tree and run number, so different realisations are independent. We estimate the uncertainty on each median from the 16th and 84th percentiles of bootstrap resamples over trees.

The precise offset at the lowest halo masses depends on how results from different assembly histories are combined. Taking the ratio of the median stellar masses over trees gives offsets of 162 and 127 per cent at $5\times10^6$ and $10^7\,\Msun$, compared with 150 and 135 per cent for the median of the per-tree ratios adopted in this work. The two estimates agree to within 1 per cent at $3\times10^7\,\Msun$ and above.

The fixed-count and fixed-delay tests and the event-size sweep use 100 trees with 10 discrete runs per tree. In the event-size sweep, the unmodified model ($k=1$) gives 135 and 14 per cent at $10^7$ and $10^8\,\Msun$, consistent with the fiducial ensemble (135 and 15 per cent). The offset remains when the number and timing of supernovae are fixed.

The variation in stellar mass between discrete realisations of the same tree is 0.087, 0.063, 0.043, 0.021 and 0.003\,dex at halo masses of $5\times10^6$, $10^7$, $3\times10^7$, $10^8$ and $10^9\,\Msun$, respectively. This variation is small compared with the offset between the discrete and continuous treatments.

% ============================================================
\section{Metal enrichment}
\label{sec:metals}
% ============================================================

\begin{figure}
\centering
\includegraphics[width=3.3in,trim=239.61bp 0 0 14bp,clip]{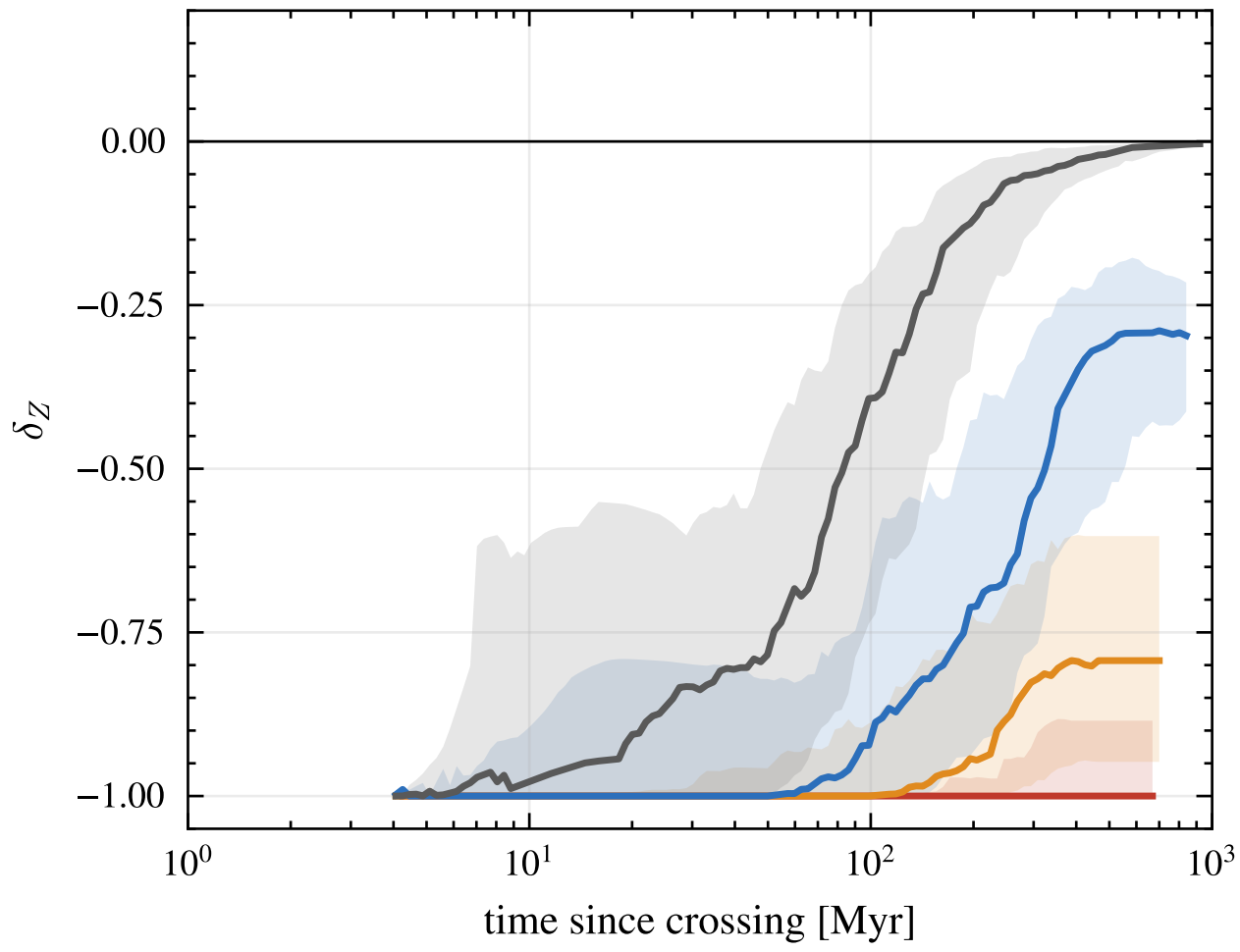}
\caption{Offset in stellar metallicity, $\delta_Z=\langle\Zs\rangle/\Zs^{\rm cont}-1$, against time since star formation began, for the same trees as Figure~\ref{fig:offsettime}.
Colours show $10^7$ (red), $3\times10^7$ (orange), $10^8$ (blue) and $10^9\,\Msun$ (grey). A value of $\delta_Z=-1$ means that the discrete runs form stars of zero metallicity.}
\label{fig:metaloffset}
\end{figure}

With pristine inflow, a supernova can remove the enriched gas from the reservoir, after which subsequent stars form from pristine material. During the first 50\,Myr, the stellar metallicity in the discrete runs is zero at $10^7$--$10^8\,\Msun$ (Figure~\ref{fig:metaloffset}). By $z=5$, the stellar-metallicity offsets are $-100$, $-72$ and $-30$ per cent at $10^7$, $3\times10^7$ and $10^8\,\Msun$, respectively, and are within 0.2 per cent of zero at $10^9\,\Msun$. At $10^9\,\Msun$ the offset is large at early times and returns to zero as the stellar masses converge.

The fraction of discrete runs in which no stars contain supernova metals is 98, 67, 6, 0 and 0 per cent at $5\times10^6$, $10^7$, $3\times10^7$, $10^8$ and $10^9\,\Msun$, respectively. When supernova-driven gas removal is set to zero, all runs become enriched. This identifies gas removal as the cause of the unenriched runs, although the model does not follow the subsequent transport or mixing of the metals. This simplification is unlikely to be conservative: star-by-star simulations of Population~III enrichment find that supernova energy typically carries metals beyond the virial radius of halos with $M_{\rm h}\lesssim10^7\,\Msun$, delaying rather than preventing self-enrichment \citep{Mead2025}, and that the instantaneous, homogeneous mixing we assume holds only within about 100\,pc and 7\,Myr of a star-forming region \citep{Mead2026}. Both effects act in the same as our gas-removal mechanism, and so our one-zone model likely understates how readily discrete feedback can delay enrichment in these halos. The resulting absence of pre-enrichment also leads to metallicities that are too low for comparison with the faintest observed systems (\S\ref{sec:discussion}). We interpret this behaviour as a consequence of the gas-removal prescription and not a prediction for the metallicities of observed dwarf galaxies.

\section*{Acknowledgements}
The authors warmly thank Anand Menon for his development of the \textsc{Ashvini} codebase. SB acknowledges the support of grant PID2022-138855NB-C32 funded by MICIU/AEI/10.13039/501100011033 and ERDF/EU, and  project PPIT2024-31833, cofunded by EU-Ministerio de Hacienda y Función P\'{u}blica-Fondos Europeos-Junta de Andaluc\'{i}a-Consejería de Universidad, Investigaci\'{o}n e Innovaci\'{o}n.  We acknowledge the use of the OzSTAR national facility at Swinburne University of Technology. This research uses \textsc{Python} \citep{vanRossum.1995} libraries, including \textsc{numpy} \citep{Harris2020} and \textsc{matplotlib} \citep{Hunter2007}. Generative AI tools were used to assist with the consistency checks of the manuscript; all scientific content, analysis, interpretation and conclusions were developed and verified by the authors, who take full responsibility for the final manuscript.

\section*{Data Availability}
All data are available from the corresponding author upon request.

\bibliographystyle{aasjournal}

\bibliography{references}

@ARTICLE{Andersson2026,
       author = {{Andersson}, Eric P. and {Rey}, Martin P. and {Yates}, Robert M. and {Read}, Justin I. and {Agertz}, Oscar and {Ji}, Alexander P. and {Mead}, Jennifer and {Brauer}, Kaley and {Mac Low}, Mordecai-Mark},
        title = "{EDGE-INFERNO: How chemical enrichment assumptions impact the individual stars of a simulated ultra-faint dwarf galaxy}",
      journal = {\aap},
         year = 2026,
        month = mar,
       volume = {707},
          eid = {A112},
        pages = {A112},
          doi = {10.1051/0004-6361/202558029},
archivePrefix = {arXiv},
       eprint = {2511.05695},
 primaryClass = {astro-ph.GA},
       adsurl = {https://ui.adsabs.harvard.edu/abs/2026A&A...707A.112A}
}

@ARTICLE{Applebaum2020,
       author = {{Applebaum}, Elaad and {Brooks}, Alyson M. and {Quinn}, Thomas R. and {Christensen}, Charlotte R.},
        title = "{Roll of the dice: a stochastically sampled IMF alters the stellar content of simulated dwarf galaxies}",
      journal = {\mnras},
         year = 2020,
        month = feb,
       volume = {492},
       number = {1},
        pages = {8-21},
          doi = {10.1093/mnras/stz3331},
archivePrefix = {arXiv},
       eprint = {1811.00022},
 primaryClass = {astro-ph.GA},
       adsurl = {https://ui.adsabs.harvard.edu/abs/2020MNRAS.492....8A}
}

@ARTICLE{Brauer2025,
       author = {{Brauer}, Kaley and {Emerick}, Andrew and {Mead}, Jennifer and {Ji}, Alexander P. and {Wise}, John H. and {Bryan}, Greg L. and {Mac Low}, Mordecai-Mark and {C{\^o}t{\'e}}, Benoit and {Andersson}, Eric P. and {Frebel}, Anna},
        title = "{AEOS: Star-by-star Cosmological Simulations of Early Chemical Enrichment and Galaxy Formation}",
      journal = {\apj},
         year = 2025,
        month = feb,
       volume = {980},
       number = {1},
          eid = {41},
        pages = {41},
          doi = {10.3847/1538-4357/ada4a1},
archivePrefix = {arXiv},
       eprint = {2410.16366},
 primaryClass = {astro-ph.GA},
       adsurl = {https://ui.adsabs.harvard.edu/abs/2025ApJ...980...41B}
}

@ARTICLE{BrauerIMF2025,
       author = {{Brauer}, Kaley and {Mead}, Jennifer and {Wise}, John H. and {Bryan}, Greg L. and {Mac Low}, Mordecai-Mark and {Ji}, Alexander P. and {Emerick}, Andrew and {Andersson}, Eric P. and {Frebel}, Anna and {C{\^o}t{\'e}}, Benoit},
        title = "{AEOS: The Impact of Population III Initial Mass Function and Star-by-star Models in Galaxy Simulations}",
      journal = {\apj},
         year = 2025,
        month = nov,
       volume = {993},
       number = {2},
        pages = {2},
          doi = {10.3847/1538-4357/ae06a4},
archivePrefix = {arXiv},
       eprint = {2502.20433},
 primaryClass = {astro-ph.GA},
       adsurl = {https://ui.adsabs.harvard.edu/abs/2025ApJ...993....2B}
}

@ARTICLE{Bromm2003,
       author = {{Bromm}, Volker and {Yoshida}, Naoki and {Hernquist}, Lars},
        title = "{The First Supernova Explosions in the Universe}",
      journal = {\apjl},
         year = 2003,
        month = oct,
       volume = {596},
       number = {2},
        pages = {L135-L138},
          doi = {10.1086/379359},
archivePrefix = {arXiv},
       eprint = {astro-ph/0305333},
 primaryClass = {astro-ph},
       adsurl = {https://ui.adsabs.harvard.edu/abs/2003ApJ...596L.135B}
}

@ARTICLE{Chabrier2003,
       author = {{Chabrier}, Gilles},
        title = "{Galactic Stellar and Substellar Initial Mass Function}",
      journal = {\pasp},
         year = 2003,
       volume = {115},
        pages = {763-795},
archivePrefix = {arXiv},
       eprint = {astro-ph/0304382}
}

@ARTICLE{Cole2000,
       author = {{Cole}, Shaun and {Lacey}, Cedric G. and {Baugh}, Carlton M. and {Frenk}, Carlos S.},
        title = "{Hierarchical galaxy formation}",
      journal = {\mnras},
         year = 2000,
       volume = {319},
        pages = {168},
          doi = {10.1046/j.1365-8711.2000.03879.x},
archivePrefix = {arXiv},
       eprint = {astro-ph/0007281}
}

@ARTICLE{Croton2006,
       author = {{Croton}, Darren J. and {Springel}, Volker and {White}, Simon D.~M. and {De Lucia}, Gabriella and {Frenk}, Carlos S. and {Gao}, Liang and {Jenkins}, Adrian and {Kauffmann}, Guinevere and {Navarro}, Julio F. and {Yoshida}, Naoki},
        title = "{The many lives of active galactic nuclei: cooling flows, black holes and the luminosities and colours of galaxies}",
      journal = {\mnras},
         year = 2006,
       volume = {365},
        pages = {11-28},
          doi = {10.1111/j.1365-2966.2005.09675.x},
archivePrefix = {arXiv},
       eprint = {astro-ph/0508046}
}

@ARTICLE{DekelSilk1986,
       author = {{Dekel}, A. and {Silk}, J.},
        title = "{The Origin of Dwarf Galaxies, Cold Dark Matter, and Biased Galaxy Formation}",
      journal = {\apj},
         year = 1986,
        month = apr,
       volume = {303},
        pages = {39},
          doi = {10.1086/164050},
       adsurl = {https://ui.adsabs.harvard.edu/abs/1986ApJ...303...39D}
}

@ARTICLE{Efstathiou1992,
       author = {{Efstathiou}, G.},
        title = "{Suppressing the formation of dwarf galaxies via photoionization}",
      journal = {\mnras},
         year = 1992,
        month = may,
       volume = {256},
       number = {2},
        pages = {43P-47P},
          doi = {10.1093/mnras/256.1.43P},
       adsurl = {https://ui.adsabs.harvard.edu/abs/1992MNRAS.256P..43E}
}

@ARTICLE{Ekstrom2012,
       author = {{Ekstr{\"o}m}, S. and {Georgy}, C. and {Eggenberger}, P. and {Meynet}, G. and {Mowlavi}, N. and {Wyttenbach}, A. and {Granada}, A. and {Decressin}, T. and {Hirschi}, R. and {Frischknecht}, U. and {Charbonnel}, C. and {Maeder}, A.},
        title = "{Grids of stellar models with rotation. I. Models from 0.8 to 120 M$_\odot$ at solar metallicity (Z = 0.014)}",
      journal = {\aap},
         year = 2012,
        month = jan,
       volume = {537},
          eid = {A146},
        pages = {A146},
          doi = {10.1051/0004-6361/201117751},
archivePrefix = {arXiv},
       eprint = {1110.5049},
 primaryClass = {astro-ph.SR},
       adsurl = {https://ui.adsabs.harvard.edu/abs/2012A&A...537A.146E}
}

@ARTICLE{FaucherGiguere2018,
       author = {{Faucher-Gigu{\`e}re}, Claude-Andr{\'e}},
        title = "{A model for the origin of bursty star formation in galaxies}",
      journal = {\mnras},
         year = 2018,
        month = jan,
       volume = {473},
       number = {3},
        pages = {3717-3731},
          doi = {10.1093/mnras/stx2595},
archivePrefix = {arXiv},
       eprint = {1701.04824},
 primaryClass = {astro-ph.GA},
       adsurl = {https://ui.adsabs.harvard.edu/abs/2018MNRAS.473.3717F}
}

@ARTICLE{Fielding2018,
       author = {{Fielding}, Drummond and {Quataert}, Eliot and {Martizzi}, Davide},
        title = "{Clustered supernovae drive powerful galactic winds after super-bubble breakout}",
      journal = {Monthly Notices of the Royal Astronomical Society},
         year = 2018,
       volume = {481},
        pages = {3325-3347},
       eprint = {1807.08758},
 archivePrefix = {arXiv}
}

@ARTICLE{Gentry2017,
       author = {{Gentry}, Eric S. and {Krumholz}, Mark R. and {Dekel}, Avishai and {Madau}, Piero},
        title = "{Enhanced momentum feedback from clustered supernovae}",
      journal = {Monthly Notices of the Royal Astronomical Society},
         year = 2017,
       volume = {465},
        pages = {2471-2488},
       eprint = {1606.01242},
 archivePrefix = {arXiv}
}

@ARTICLE{Georgy2013,
       author = {{Georgy}, C. and {Ekstr{\"o}m}, S. and {Eggenberger}, P. and {Meynet}, G. and {Haemmerl{\'e}}, L. and {Maeder}, A. and {Granada}, A. and {Groh}, J.~H. and {Hirschi}, R. and {Mowlavi}, N. and {Yusof}, N. and {Charbonnel}, C. and {Decressin}, T. and {Barblan}, F.},
        title = "{Grids of stellar models with rotation. III. Models from 0.8 to 120 M$_{☉}$ at a metallicity Z = 0.002}",
      journal = {\aap},
         year = 2013,
        month = oct,
       volume = {558},
          eid = {A103},
        pages = {A103},
          doi = {10.1051/0004-6361/201322178},
archivePrefix = {arXiv},
       eprint = {1308.2914},
 primaryClass = {astro-ph.SR},
       adsurl = {https://ui.adsabs.harvard.edu/abs/2013A&A...558A.103G}
}

@ARTICLE{Grudic2023,
       author = {{Grudi{\'c}}, Michael Y. and {Offner}, Stella S. R. and {Guszejnov}, D{\'a}vid and {Faucher-Gigu{\`e}re}, Claude-Andr{\'e} and {Hopkins}, Philip F.},
        title = "{Does God play dice with star clusters?}",
      journal = {The Open Journal of Astrophysics},
         year = 2023,
        month = dec,
       volume = {6},
          eid = {48},
          doi = {10.21105/astro.2307.00052},
       eprint = {2307.00052},
 archivePrefix = {arXiv},
       adsurl = {https://ui.adsabs.harvard.edu/abs/2023OJAp....6E..48G}
}

@ARTICLE{GrudicHopkins2019,
       author = {{Grudi{\'c}}, Michael Y. and {Hopkins}, Philip F.},
        title = "{The elephant in the room: the importance of the details of massive star formation in molecular clouds}",
      journal = {\mnras},
         year = 2019,
        month = sep,
       volume = {488},
        pages = {2970-2975},
          doi = {10.1093/mnras/stz1820},
       eprint = {1809.08344},
 archivePrefix = {arXiv},
       adsurl = {https://ui.adsabs.harvard.edu/abs/2019MNRAS.488.2970G}
}

@ARTICLE{Gutcke2021,
       author = {{Gutcke}, Thales A. and {Pakmor}, R{\"u}diger and {Naab}, Thorsten and {Springel}, Volker},
        title = "{LYRA - I. Simulating the multiphase ISM of a dwarf galaxy with variable energy supernovae from individual stars}",
      journal = {\mnras},
         year = 2021,
        month = mar,
       volume = {501},
       number = {4},
        pages = {5597-5615},
          doi = {10.1093/mnras/staa3875},
archivePrefix = {arXiv},
       eprint = {2010.07311},
 primaryClass = {astro-ph.GA},
       adsurl = {https://ui.adsabs.harvard.edu/abs/2021MNRAS.501.5597G}
}

@Article{Harris2020,
  author    = {Harris, Charles R. and Millman, K. Jarrod and van der Walt, Stéfan J. and Gommers, Ralf and Virtanen, Pauli and Cournapeau, David and Wieser, Eric and Taylor, Julian and Berg, Sebastian and Smith, Nathaniel J. and Kern, Robert and Picus, Matti and Hoyer, Stephan and van Kerkwijk, Marten H. and Brett, Matthew and Haldane, Allan and del Río, Jaime Fernández and Wiebe, Mark and Peterson, Pearu and Gérard-Marchant, Pierre and Sheppard, Kevin and Reddy, Tyler and Weckesser, Warren and Abbasi, Hameer and Gohlke, Christoph and Oliphant, Travis E.},
  journal   = {Nature},
  title     = {Array programming with NumPy},
  year      = {2020},
  issn      = {1476-4687},
  month     = sep,
  number    = {7825},
  pages     = {357--362},
  volume    = {585},
  doi       = {10.1038/s41586-020-2649-2},
  publisher = {Springer Science and Business Media LLC},
}

@ARTICLE{Henriques2020,
       author = {{Henriques}, Bruno M.~B. and {Yates}, Robert M. and {Fu}, Jian and {Guo}, Qi and {Kauffmann}, Guinevere and {Srisawat}, Chaichalit and {Thomas}, Peter A. and {White}, Simon D.~M.},
        title = "{L-GALAXIES 2020: Spatially resolved cold gas phases, star formation, and chemical enrichment in galactic discs}",
      journal = {\mnras},
         year = 2020,
       volume = {491},
       number = {4},
        pages = {5795-5814},
          doi = {10.1093/mnras/stz3233},
archivePrefix = {arXiv},
       eprint = {2003.05944}
}

@ARTICLE{Hirai2021,
       author = {{Hirai}, Yutaka and {Fujii}, Michiko S. and {Saitoh}, Takayuki R.},
        title = "{SIRIUS project. I. Star formation models for star-by-star simulations of star clusters and galaxy formation}",
      journal = {\pasj},
         year = 2021,
        month = aug,
       volume = {73},
        pages = {1036-1056},
          doi = {10.1093/pasj/psab038},
       eprint = {2005.12906},
 archivePrefix = {arXiv},
       adsurl = {https://ui.adsabs.harvard.edu/abs/2021PASJ...73.1036H}
}

@ARTICLE{Hunter2007,
  author={Hunter, John D.},
  journal={Computing in Science \& Engineering}, 
  title={Matplotlib: A 2D Graphics Environment}, 
  year={2007},
  volume={9},
  number={3},
  pages={90-95},
  doi={10.1109/MCSE.2007.55}
}

@ARTICLE{Hutter2021,
       author = {{Hutter}, Anne and {Dayal}, Pratika and {Yepes}, Gustavo and {Gottl{\"o}ber}, Stefan and {Legrand}, Laurent and {Ucci}, Graziano},
        title = "{Astraeus I: The interplay between galaxy formation and reionization}",
      journal = {\mnras},
         year = 2021,
       volume = {503},
       number = {3},
        pages = {3698},
          doi = {10.1093/mnras/stab602},
archivePrefix = {arXiv},
       eprint = {2004.08401}
}

@ARTICLE{JeonGo2026,
       author = {{Jeon}, Myoungwon and {Go}, Minsung},
        title = "{Exploring effects of IMF sampling and SN feedback injection on star formation and metallicity in ultra-faint dwarf galaxies}",
      journal = {\mnras},
         year = 2026,
        month = jan,
       volume = {545},
       number = {3},
          eid = {staf2158},
        pages = {staf2158},
          doi = {10.1093/mnras/staf2158},
archivePrefix = {arXiv},
       eprint = {2411.17862},
 primaryClass = {astro-ph.GA},
       adsurl = {https://ui.adsabs.harvard.edu/abs/2026MNRAS.545f2158J}
}

@ARTICLE{KellerKruijssen2022,
       author = {{Keller}, Benjamin W. and {Kruijssen}, J.~M. Diederik},
        title = "{Uncertainties in supernova input rates drive qualitative differences in simulations of galaxy evolution}",
      journal = {\mnras},
         year = 2022,
       volume = {512},
       number = {1},
        pages = {199-215},
          doi = {10.1093/mnras/stac511}
}

@ARTICLE{Kitayama2004,
       author = {{Kitayama}, Tetsu and {Yoshida}, Naoki and {Susa}, Hajime and {Umemura}, Masayuki},
        title = "{The Structure and Evolution of Early Cosmological H II Regions}",
      journal = {The Astrophysical Journal},
         year = 2004,
       volume = {613},
        pages = {631-645},
       eprint = {astro-ph/0406280},
 archivePrefix = {arXiv}
}

@ARTICLE{KitayamaYoshida2005,
       author = {{Kitayama}, Tetsu and {Yoshida}, Naoki},
        title = "{Supernova Explosions in the Early Universe: Evolution of Radiative Remnants and the Halo Destruction Efficiency}",
      journal = {\apj},
         year = 2005,
        month = sep,
       volume = {630},
       number = {2},
        pages = {675-688},
          doi = {10.1086/432114},
archivePrefix = {arXiv},
       eprint = {astro-ph/0505368},
 primaryClass = {astro-ph},
       adsurl = {https://ui.adsabs.harvard.edu/abs/2005ApJ...630..675K}
}

@MISC{KravtsovBelokurov2024,
       author = {{Kravtsov}, Andrey and {Belokurov}, Vasily},
        title = "{Stochastic star formation and the abundance of $z>10$ UV-bright galaxies}",
         year = 2024,
       eprint = {2405.04578},
 archivePrefix = {arXiv}
}

@ARTICLE{Lenoble2024,
       author = {{Lenoble}, R. and {Commer{\c{c}}on}, B. and {Rosdahl}, J.},
        title = "{Simulations of early structure formation: Properties of halos that host primordial star formation}",
      journal = {\aap},
         year = 2024,
        month = may,
       volume = {685},
          eid = {A7},
        pages = {A7},
          doi = {10.1051/0004-6361/202347918},
archivePrefix = {arXiv},
       eprint = {2401.16821},
 primaryClass = {astro-ph.GA},
       adsurl = {https://ui.adsabs.harvard.edu/abs/2024A&A...685A...7L}
}

@ARTICLE{Mead2025,
       author = {{Mead}, Jennifer and {Brauer}, Kaley and {Bryan}, Greg L. and {Mac Low}, Mordecai-Mark and {Ji}, Alexander P. and {Wise}, John H. and {Emerick}, Andrew and {Andersson}, Eric P. and {Frebel}, Anna and {C{\^o}t{\'e}}, Benoit},
        title = "{AEOS: Transport of Metals from Minihalos following Population III Stellar Feedback}",
      journal = {\apj},
         year = 2025,
        month = feb,
       volume = {980},
       number = {1},
        pages = {62},
          doi = {10.3847/1538-4357/ada3c1},
archivePrefix = {arXiv},
       eprint = {2411.14209},
 primaryClass = {astro-ph.GA},
       adsurl = {https://ui.adsabs.harvard.edu/abs/2025ApJ...980...62M}
}

@ARTICLE{Mead2026,
       author = {{Mead}, Jennifer and {Brauer}, Kaley and {Bryan}, Greg L. and {Mac Low}, Mordecai-Mark and {Ji}, Alexander P. and {Wise}, John H. and {Andersson}, Eric P. and {Frebel}, Anna and {Emerick}, Andrew and {C{\^o}t{\'e}}, Benoit},
        title = "{AEOS Is Mixing It Up: The (In)homogeneity of Metal Mixing Following Population III Star Formation}",
      journal = {\apj},
         year = 2026,
        month = may,
       volume = {1002},
       number = {1},
        pages = {31},
          doi = {10.3847/1538-4357/ae5953},
archivePrefix = {arXiv},
       eprint = {2509.13580},
 primaryClass = {astro-ph.GA},
       adsurl = {https://ui.adsabs.harvard.edu/abs/2026ApJ.1002...31M}
}

@ARTICLE{MenonBaluPower2026,
       author = {{Menon}, Anand and {Balu}, Sreedhar and {Power}, Chris},
        title = "{On bursty star formation during cosmological reionisation ─ Influence on the metal and dust content of low-mass galaxies}",
      journal = {\pasa},
         year = 2026,
        month = mar,
       volume = {43},
          eid = {e050},
        pages = {e050},
          doi = {10.1017/pasa.2026.10182},
archivePrefix = {arXiv},
       eprint = {2508.08363},
 primaryClass = {astro-ph.GA},
       adsurl = {https://ui.adsabs.harvard.edu/abs/2026PASA...43...50M}
}

@ARTICLE{MenonPower2024,
       author = {{Menon}, Anand and {Power}, Chris},
        title = "{On bursty star formation during cosmological reionisation ─ how does it influence the baryon mass content of dark matter halos?}",
      journal = {\pasa},
         year = 2024,
        month = may,
       volume = {41},
          eid = {e049},
        pages = {e049},
          doi = {10.1017/pasa.2024.39},
archivePrefix = {arXiv},
       eprint = {2405.03211},
 primaryClass = {astro-ph.GA},
       adsurl = {https://ui.adsabs.harvard.edu/abs/2024PASA...41...49M}
}

@ARTICLE{Mutch2016,
       author = {{Mutch}, Simon J. and {Geil}, Paul M. and {Poole}, Gregory B. and {Angel}, Paul W. and {Duffy}, Alan R. and {Mesinger}, Andrei and {Wyithe}, J. Stuart B.},
        title = "{Dark-ages reionization and galaxy formation simulation - III. Modelling galaxy formation and the epoch of reionization}",
      journal = {\mnras},
         year = 2016,
       volume = {462},
       number = {1},
        pages = {250-276},
          doi = {10.1093/mnras/stw1506},
archivePrefix = {arXiv},
       eprint = {1512.00562}
}

@ARTICLE{Okamoto2008,
       author = {{Okamoto}, Takashi and {Gao}, Liang and {Theuns}, Tom},
        title = "{Mass loss of galaxies due to an ultraviolet background}",
      journal = {\mnras},
         year = 2008,
       volume = {390},
       number = {3},
        pages = {920-928},
          doi = {10.1111/j.1365-2966.2008.13830.x},
archivePrefix = {arXiv},
       eprint = {0806.0378}
}

@MISC{PanKravtsov2023,
       author = {{Pan}, Yue and {Kravtsov}, Andrey},
        title = "{Modelling Stochastic Star Formation History of Dwarf Galaxies in GRUMPY}",
         year = 2023,
       eprint = {2310.08636},
 archivePrefix = {arXiv}
}

@ARTICLE{Parkinson2008,
       author = {{Parkinson}, Hannah and {Cole}, Shaun and {Helly}, John},
        title = "{Generating dark matter halo merger trees}",
      journal = {\mnras},
         year = 2008,
        month = jan,
       volume = {383},
       number = {2},
        pages = {557-564},
          doi = {10.1111/j.1365-2966.2007.12517.x},
archivePrefix = {arXiv},
       eprint = {0708.1382},
 primaryClass = {astro-ph},
       adsurl = {https://ui.adsabs.harvard.edu/abs/2008MNRAS.383..557P}
}

@ARTICLE{PowerBalu2026,
       author = {{Power}, C. and {Balu}, S.},
        title = "{foraois: Fast, Flexible Extended Press-Schechter Merger Trees with a General First-Crossing Solver}",
      journal = {arXiv e-prints},
         year = 2026,
        month = nov,
          eid = {arXiv:2611.01580},
        pages = {arXiv:2611.01580},
          doi = {10.48550/arXiv.2611.01580},
archivePrefix = {arXiv},
       eprint = {2611.01580},
 primaryClass = {astro-ph.CO},
       adsurl = {https://ui.adsabs.harvard.edu/abs/2026arXiv261101580P}
}

@ARTICLE{Sanati2023,
       author = {{Sanati}, Mahsa and {Jeanquartier}, Fabien and {Revaz}, Yves and {Jablonka}, Pascale},
        title = "{How much metal did the first stars provide to the ultra-faint dwarfs?}",
      journal = {\aap},
         year = 2023,
        month = jan,
       volume = {669},
          eid = {A94},
        pages = {A94},
          doi = {10.1051/0004-6361/202244309},
archivePrefix = {arXiv},
       eprint = {2206.11351},
 primaryClass = {astro-ph.GA},
       adsurl = {https://ui.adsabs.harvard.edu/abs/2023A&A...669A..94S}
}

@ARTICLE{Smith2021,
       author = {{Smith}, Matthew C. and {Bryan}, Greg L. and {Somerville}, Rachel S. and {Hu}, Chia-Yu and {Teyssier}, Romain and {Burkhart}, Blakesley and {Hernquist}, Lars},
        title = "{Efficient early stellar feedback can suppress galactic outflows by reducing supernova clustering}",
      journal = {\mnras},
         year = 2021,
        month = sep,
       volume = {506},
        pages = {3882-3915},
          doi = {10.1093/mnras/stab1896},
       eprint = {2009.11309},
 archivePrefix = {arXiv},
       adsurl = {https://ui.adsabs.harvard.edu/abs/2021MNRAS.506.3882S}
}

@ARTICLE{Smith2021IMF,
       author = {{Smith}, Matthew C.},
        title = "{The sensitivity of stellar feedback to IMF averaging versus IMF sampling in galaxy formation simulations}",
      journal = {\mnras},
         year = 2021,
       volume = {502},
        pages = {5417},
          doi = {10.1093/mnras/stab291},
       eprint = {2010.10533},
 archivePrefix = {arXiv}
}

@ARTICLE{Smith2026,
       author = {{Smith}, Matthew C.},
        title = "{Imladris: a detailed and flexible model for galaxy simulations with individual stars}",
      journal = {\mnras},
         year = 2026,
        month = jul,
       volume = {549},
       number = {3},
          eid = {stag959},
        pages = {stag959},
          doi = {10.1093/mnras/stag959},
archivePrefix = {arXiv},
       eprint = {2604.00100},
 primaryClass = {astro-ph.GA},
       adsurl = {https://ui.adsabs.harvard.edu/abs/2026MNRAS.549ag959S}
}

@ARTICLE{SpringelHernquist2003,
       author = {{Springel}, Volker and {Hernquist}, Lars},
        title = "{Cosmological smoothed particle hydrodynamics simulations: a hybrid multiphase model for star formation}",
      journal = {\mnras},
         year = 2003,
       volume = {339},
       number = {2},
        pages = {289-311},
          doi = {10.1046/j.1365-8711.2003.06206.x},
archivePrefix = {arXiv},
       eprint = {astro-ph/0206393}
}

@ARTICLE{Su2018,
       author = {{Su}, Kung-Yi and {Hopkins}, Philip F. and {Hayward}, Christopher C. and {Ma}, Xiangcheng and {Boylan-Kolchin}, Michael and {Kasen}, Daniel and {Kere{\v{s}}}, Du{\v{s}}an and {Faucher-Gigu{\`e}re}, Claude-Andr{\'e} and {Orr}, Matthew E. and {Wheeler}, Coral},
        title = "{Discrete effects in stellar feedback: Individual Supernovae, Hypernovae, and IMF Sampling in Dwarf Galaxies}",
      journal = {\mnras},
         year = 2018,
        month = oct,
       volume = {480},
       number = {2},
        pages = {1666-1675},
          doi = {10.1093/mnras/sty1928},
archivePrefix = {arXiv},
       eprint = {1712.02795},
 primaryClass = {astro-ph.GA},
       adsurl = {https://ui.adsabs.harvard.edu/abs/2018MNRAS.480.1666S}
}

@manual{vanRossum.1995,
 title                = {Python Reference Manual},
 author               = {van Rossum, Guido},
 number               = {R 9525},
 year                 = 1995,
 month                = jan,
 organization            = {Centrum voor Wiskunde en Informatica},
 note                 = {}
}

@ARTICLE{Whalen2008,
       author = {{Whalen}, Daniel and {van Veelen}, Bob and {O'Shea}, Brian W. and {Norman}, Michael L.},
        title = "{The destruction of cosmological minihalos by primordial supernovae}",
      journal = {The Astrophysical Journal},
         year = 2008,
       volume = {682},
        pages = {49-67},
          doi = {10.1086/589643}
}

@ARTICLE{Zhang2025,
       author = {{Zhang}, Eric and {Sales}, Laura V. and {Gutcke}, Thales A. and {Deng}, Yunwei and {Li}, Hui and {Pakmor}, R{\"u}diger and {Marinacci}, Federico and {Springel}, Volker and {Vogelsberger}, Mark and {Torrey}, Paul and {Liu}, Boyuan and {Kannan}, Rahul and {Smith}, Aaron and {Bryan}, Greg L.},
        title = "{The Entangling of Supernova Feedback Impacts with Coarsening Simulation Resolution}",
      journal = {arXiv e-prints},
         year = 2025,
        month = oct,
          eid = {arXiv:2510.02432},
        pages = {arXiv:2510.02432},
          doi = {10.48550/arXiv.2510.02432},
archivePrefix = {arXiv},
       eprint = {2510.02432},
 primaryClass = {astro-ph.GA},
       adsurl = {https://ui.adsabs.harvard.edu/abs/2025arXiv251002432Z}
}

\end{document}